\documentclass[aps,preprint,nofootinbib,floatfix,a4paper,prd]{revtex4-1}
\usepackage{cmap}
\usepackage[T1]{fontenc}
\usepackage{titlesec}
\usepackage{graphicx}
\usepackage{caption}
\usepackage{subcaption}
\usepackage{color}
\usepackage[latin1]{inputenc}
\usepackage{amsmath,amssymb}
\usepackage{hyperref}
\usepackage{epstopdf}
\usepackage{geometry}    

\def\be{\begin{equation}}
\def\ee{\end{equation}}
\def\bea{\begin{eqnarray}}
\def\eea{\end{eqnarray}}

\newcommand{\lsim}{\mathrel{\mathop{\kern 0pt \rlap
  {\raise.2ex\hbox{$<$}}}
  \lower.9ex\hbox{\kern-.190em $\sim$}}}
\newcommand{\gsim}{\mathrel{\mathop{\kern 0pt \rlap
  {\raise.2ex\hbox{$>$}}}
  \lower.9ex\hbox{\kern-.190em $\sim$}}}

\def  \bcen   {\begin{center}}
\def  \ecen   {\end{center}}
\def  \beq    {\begin{equation}}
\def  \eeq    {\end{equation}}
\def  \bpm    {\begin{pmatrix}}
\def  \epm    {\end{pmatrix}}
\def  \beqa   {\begin{eqnarray}}
\def  \eeqa   {\end{eqnarray}}
\def  \nn     {\nonumber }
\def\bea{\begin{eqnarray}}
\def\eea{\end{eqnarray}}

\def\ga   {\gamma}

\def\la   {\lambda}

\def\nn{\nonumber}
\def\lee { \left( }
\def\rii { \right) }

\def\De {\Delta}

\def\to {\rightarrow}

\begin{document}

\title{Pair Production of Higgs Boson in G2HDM at the LHC}

\author{Chuan-Ren Chen$^{1}$, Yu-Xiang Lin$^{1}$, Van Que Tran$^{1}$ and Tzu-Chiang Yuan$^{2}$}

\affiliation{$^1$Department of Physics, National Taiwan Normal University, Taipei 116, Taiwan \\
$^2$Institute of Physics, Academia Sinica, Nangang, Taipei 11529, Taiwan}

\begin{abstract}
Pair production of Higgs boson at the Large Hadron Collider (LHC)  is known to be important for the determination of Higgs boson self-coupling and the probe of new physics beyond the Standard Model (SM), especially the existence of new fundamental scalar boson. In this paper we study in detail the Higgs pair production at the LHC in a well-motivated model, the Gauged Two Higgs Doublet Model (G2HDM) in which the two Higgs doublets are properly embedded into a gauged $SU(2)_H$ and a dark matter candidate emerges naturally due to the gauge symmetry. Besides the deviations of Higgs couplings from the SM predictions, the existence of new scalars could enhance the production cross section of Higgs boson pair at the LHC significantly. However, when we take into account the relic density of dark matter and the null result in its direct search, only moderate enhancement can be maintained. We also comment on the capability of distinguishing the signal of a new generic scalar from the SM at the LHC, assuming the Higgs pair production cross sections are the same.  
\end{abstract}

\maketitle

\section{Introduction \label{intro}}
This year marks the fiftieth anniversary of the Standard Model (SM) of particle physics since its inception in 1967~\cite{SM@50}.
Higgs boson, the long sought last particle responsible for spontaneously symmetry breaking
via the Higgs mechanism in SM, was ultimately discovered in 2012 at the Large Hadron Collider (LHC)
with a relatively light mass of 125 GeV.
Since then many efforts were made to determine whether the observed Higgs boson 
is indeed the one predicted in SM. 
In particular all the measurements related to the Higgs couplings with gauge bosons had been 
analyzed in details and no deviations from SM predictions were found.
Recently, six years after its discovery, the Higgs boson decaying into its dominant mode, a pair of bottom quarks, 
has finally been observed at a significance above 5 standard deviations 
and also in line with the SM expectation~\cite{Aaboud:2018zhk,Sirunyan:2018kst}.

Despite its many triumphs, SM leaves us with many questions to be answered. For instance, just to name a few, 
the origin of flavor remains to be a puzzle, the hierarchy/fine tuning problem, and so on.
Even for the Higgs sector, the self-coupling of the Higgs boson and the related issue of the 
shape of the Higgs potential are needed to be determined.  Moreover the observations of neutrino oscillations, 
relic density of dark matter (DM), matter-antimatter asymmetry {\it etc.} imply there must be new physics beyond the SM. 

Many new models had been proposed in the literature to address the above issues, 
and one of the simplest proposals is extending the scalar sector of the SM. 
Along this direction,  the general two Higgs doublet model (2HDM) is a simple extension~\cite{Branco:2011iw} 
by just adding one more Higgs  doublet to the SM.
One particular type of 2HDM is the inert Higgs doublet model 
(IHDM)~\cite{Deshpande:1977rw,Ma:2006km,Barbieri:2006dq,LopezHonorez:2006gr,Arhrib:2013ela} 
in which the neutral component of the second Higgs doublet is a DM candidate. The stability of this DM candidate 
is ensured by imposing a discrete $Z_2$ symmetry on the scalar potential of the model. 

Recently, a Gauged Two Higgs Doublet model (G2HDM)~\cite{Huang:2015wts} was proposed.  
Two Higgs doublets $H_1$ and $H_2$ are introduced and gauged under a non-abelian $SU(2)_H$ in the model. 
The neutral component of $H_2$ is stable under the protection of $SU(2)_H$ gauge symmetry and hence can be a DM candidate.
To make the model self-consistent, however, more new particles are introduced in G2HDM, including an additional $SU(2)_H$ doublet, a $SU(2)_H$ triplet, and heavy $SU(2)_L$ singlet Dirac fermions. 
Some phenomenology of G2HDM at the LHC had been explored previously in~\cite{Huang:2015wts,Huang:2015rkj} 
for Higgs physics and in~\cite{Huang:2017bto} for the new gauge bosons. Recently, a detailed study of the
theoretical and Higgs phenomenological constraints in G2HDM has been presented in \cite{Arhrib:2018sbz}.
In this paper, we focus on the Higgs pair production in G2HDM at the LHC, which plays a crucial role in determination of Higgs boson self-coupling.

In the SM, after electroweak symmetry breaking the scalar potential can be written as
\be
V_{\rm SM} = \frac{m_h^2}{2} h^2 + \lambda_{\rm SM} v h^3 + \frac{\lambda_{\rm SM}}{4} h^4,
\ee
where $ \lambda_{\rm SM} = \frac{m_h^2}{2v^2}$ with $m_h$ being the Higgs boson mass and $v = 246$ GeV. 
While the first term in the above Higgs potential (or Higgs boson mass term) has been measured at the LHC, the second and third terms (Higgs boson self-coupling terms) have not yet been measured. These self-couplings are key parameters for the reconstruction of the Higgs potential that tells us how the electroweak symmetry breaking really happens and whether the Higgs sector agrees with the SM.  In the G2HDM, the production of Higgs pair will be affected, as compared to the SM prediction, in a number of manners listed as follows:
\begin{itemize}
\item Modified Yukawa couplings.
\item Modified trilinear Higgs self-coupling.
\item Presence of new colored particles which can flow inside the triangle and box loops.
\item Presence of new heavy scalars which can decay into a pair of 125 GeV Higgs bosons via new trilinear scalar couplings.
This has important impact to the resonant Higgs boson pair production cross section.
\end{itemize} 
We will include all these new features in our analysis of Higgs pair production in G2HDM at the LHC.

This paper is organized as follows. In Section~\ref{model} we first review the setup of G2HDM with special focus on the scalar sector.  In Section~\ref{constraint} we review the constraints on this model that have been studied in the literature. 
In Section~\ref{higgspair} we discuss the Higgs pair production cross section in G2HDM. 
In Section~\ref{NR}, we present our numerical analysis. 
Finally, we conclude in Section~\ref{concl}.

\section{Model \label{model}}
To make this paper somewhat self-contained, we will briefly review the G2HDM in this section and refer our readers to Refs.~\cite{Huang:2015wts,Arhrib:2018sbz} for more details. The gauge groups in the G2HDM consist of $SU(3)_C \times SU(2)_L \times U(1)_Y  \times SU(2)_H \times U(1)_X$. The scalar sector includes not only the two $SU(2)_L$ Higgs doublets $H_1$, $H_2$ which form a doublet $H=(H_1,H_2)^T$ in the $SU(2)_H$ gauge group, but also a triplet $\Delta_H$ and a doublet $\Phi_H$ of this new gauge group. Note that, $\Delta_H$ and $\Phi_H$ are both singlets under the SM gauge group. Furthermore, $H$ and $\Phi_H$ are assigned to carry an additional $U(1)_X$ charge. For the fermion sector, the SM left-handed $SU(2)_L$ doublets are singlets under $SU(2)_H$, while the SM right-handed $SU(2)_L$ singlets are now paired up with new right-handed singlets to form doublets under $SU(2)_H$. Furthermore, for anomaly cancellations, new heavy left-handed fermions 
are needed, which are singlets under both $SU(2)_L$ and $SU(2)_H$ gauge groups. 
In Table ~\ref {tab:quantumnos}, we summarize the matter content and their quantum number assignments in G2HDM.
\begin{table}[htbp!]
\begin{tabular}{|c|c|c|c|c|c|}
\hline
Matter Fields & $SU(3)_C$ & $SU(2)_L$ & $SU(2)_H$ & $U(1)_Y$ & $U(1)_X$ \\
\hline \hline
$H=\left( H_1 \;\; H_2 \right)^{\rm T}$ & 1 & 2 & 2 & 1/2 & $1$ \\
$\Delta_H$ & 1 & 1 & 3 & 0 & 0 \\
$\Phi_H$ & 1 & 1 & 2 & 0 & $1$ \\
\hline\hline
$Q_L=\left( u_L \;\; d_L \right)^{\rm T}$ & 3 & 2 & 1 & 1/6 & 0\\
$U_R=\left( u_R \;\; u^H_R \right)^{\rm T}$ & 3 & 1 & 2 & 2/3 & $1$ \\
$D_R=\left( d^H_R \;\; d_R \right)^{\rm T}$ & 3 & 1 & 2 & $-1/3$ & $-1$ \\
\hline
$u_L^H$ & 3 & 1 & 1 & 2/3 & 0 \\
$d_L^H$ & 3 & 1 & 1 & $-1/3$ & 0 \\
\hline
$L_L=\left( \nu_L \;\; e_L \right)^{\rm T}$ & 1 & 2 & 1 & $-1/2$ & 0 \\
$N_R=\left( \nu_R \;\; \nu^H_R \right)^{\rm T}$ & 1 & 1 & 2 & 0 & $1$ \\
$E_R=\left( e^H_R \;\; e_R \right)^{\rm T}$ & 1 & 1 & 2 &  $-1$  &  $-1$ \\
\hline
$\nu_L^H$ & 1 & 1 & 1 & 0 & 0 \\
$e_L^H$ & 1 & 1 & 1 & $-1$ & 0 \\
\hline
\end{tabular}
\caption{Matter content and their quantum number assignments in G2HDM. 
}
\label{tab:quantumnos}
\end{table}

\subsection{ Higgs Potential \label{subsec5.1.3}}
The most general Higgs potential invariant under both $SU(2)_L\times U(1)_Y$ and $SU(2)_H \times U(1)_X$ 
is given by~\cite{Arhrib:2018sbz}
\begin{align}
V_T = V (H) + V (\Phi_H ) + V ( \De_H ) + V_{\rm mix} \left( H , \Delta_H, \Phi_H \right) \; , 
\label{eq:higgs_pot} 
\end{align}
where
\begin{align}
\label{VH1H2}
V (H) 
=&  \;  \mu^2_H   \left( H^\dag_1 H_1 + H^\dag_2 H_2 \right) 
+ \la_H   \left( H^\dag_1 H_1 + H^\dag_2 H_2 \right)^2 \nn \\
& + \la'_H \left( - H^\dag_1 H_1 H^\dag_2 H_2 
                  + H^\dag_1 H_2 H^\dag_2 H_1 \right)  \; , 
\end{align}
\begin{align}
\label{VPhi}
V ( \Phi_H ) =& \;  \mu^2_{\Phi}   \Phi_H^\dag \Phi_H  + \la_\Phi \lee \Phi_H^\dag \Phi_H  \rii^2  \; , 
 \end{align}
\begin{align}
 \label{VDelta}
V ( \De_H ) =& \; - \mu^2_{\De} {\rm Tr} \lee \De^2_H  \rii  \;  + \la_\De \lee {\rm Tr} \lee \De^2_H  \rii \rii^2 \; ,  
\end{align}
where
 \begin{align}
\De_H=
  \begin{pmatrix}
    \De_3/2   &  \De_p / \sqrt{2}  \\
    \De_m / \sqrt{2} & - \De_3/2   \\
  \end{pmatrix} = \De_H^\dagger \; {\rm with}
  \;\; \Delta_m = \left( \Delta_p \right)^* \; {\rm and} \; \left( \Delta_3 \right)^* = \Delta_3 \;    ;
 \end{align}
and the last term%
\begin{align}
\label{VMix}
V_{\rm{mix}} \left( H , \Delta_H, \Phi_H \right) = 
& \; + M_{H\De}  \lee H^\dag \De_H H \rii -  M_{\Phi\De}  \lee \Phi_H^\dag \De_H \Phi_H \rii  \nn \\
& \; + \la_{H\Phi} \lee H^\dag H  \rii  \lee \Phi_H^\dag \Phi_H \rii  
 + \la^\prime_{H\Phi} \lee H^\dag \Phi_H  \rii  \lee \Phi_H^\dag H \rii
\nn\\
& \;  +  \la_{H\De} \lee H^\dag H  \rii    {\rm Tr} \lee \De^2_H  \rii  
+ \la_{\Phi\De} \lee \Phi_H^\dag \Phi_H \rii {\rm Tr} \lee \De^2_H \rii  \; . 
\end{align}

Note also that the scalar potential in G2HDM is CP-conserving due to the fact that all terms in $V(H)$, $V(\Phi_H)$, $V(\De_H)$ and $V_{\rm mix}(H, \De_H, \Phi_H)$ are Hermitian, 
implying all the coefficients are necessarily real.

\subsection{Spontaneous Symmetry Breaking and Scalar Mass Spectrum \label{subsec5.1.4}}
\subsubsection{Spontaneous Symmetry Breaking \label{subsub5.1.3.1}}

First, let us parameterize the fields as follows 
\begin{eqnarray}
\label{HiggsComponents}
H_1 = 
\begin{pmatrix}
G^+ \\ \frac{v + h}{\sqrt 2} + i \frac{G^0}{\sqrt 2}
\end{pmatrix}
, \;
H_2 = 
\begin{pmatrix}
H^+ \\ H_2^0
\end{pmatrix}
, \;
\Phi_H = 
\begin{pmatrix}
G_H^p \\ \frac{v_\Phi + \phi_2}{\sqrt 2} + i \frac{G_H^0}{\sqrt 2}
\end{pmatrix}
, \;
\Delta_H =
\begin{pmatrix}
\frac{-v_\De + \delta_3}{2} & \frac{1}{\sqrt 2}\De_p \\ 
\frac{1}{\sqrt 2}\De_m & \frac{v_\De - \delta_3}{2}
\end{pmatrix} . \nn \\
\end{eqnarray}
where $v$, $v_\Phi$ and $v_\De$ are VEVs to be determined
by minimization of the potential. The set 
$\Psi_G \equiv \{ G^0, G^+, G^0_H, G^p_H\}$ are Goldstone bosons.
Then, inserting the VEVs $v$, $v_\Phi$, $v_\De$ into the potential $V_T$ in Eq.~\eqref{eq:higgs_pot}  
leads to
\begin{eqnarray}
\label{Vvevs}
V_T(v,  v_\De , v_\Phi) & = & 
\frac{1}{4} \left[ 
\lambda_H v^4 + \lambda_\Phi v_\Phi^4 + \lambda_\De v_\De^4 + 2 \left( \mu_H^2 v^2 + \mu_\Phi^2 v_\Phi^2 - \mu_\De^2 v_\De^2 \right) \right. \nonumber \\
&& \left. - \left( M_{H\De} v^2 + M_{\Phi\De} v_\Phi^2 \right) v_\De + \lambda_{H\Phi} 
 v^2 v_\Phi^2 + \lambda_{H\De} v^2 v_\De^2 + \lambda_{\Phi\De} v_\Phi^2 v_\De^2
\right] \; . \nonumber \\
\end{eqnarray}
By minimizing the potential in Eq.~{\eqref{Vvevs}}, we obtain the following equations satisfied by the VEVs:
\begin{eqnarray}
\label{vevv}
\left( 2\lambda_H v^2 + 2 \mu_H^2 - M_{H\De} v_\De  + \lambda_{H\Phi} 
 v_\Phi^2 + \lambda_{H\De} v_\De^2 \right)
& = & 0 \; , \\
\label{vevphi}
 \left( 2\lambda_\Phi v_\Phi^2 + 2 \mu_\Phi^2 -  M_{\Phi\De} v_\De + \lambda_{H\Phi} 
 v^2 + \lambda_{\Phi\De} v_\De^2 \right)
& = &  0 \; , \\
\label{vevdelta}
4\lambda_\De v_\De^3 - 4 \mu_\De^2 v_\De - M_{H \De} v^2 - M_{\Phi \De} v_\Phi^2 
+ 2 v_\De \left( \lambda_{H\De} v^2 + \lambda_{\Phi\De} v_\Phi^2 \right) & = & 0 \; .
\end{eqnarray}
By solving this set of coupled algebraic equations, one can get solutions for all the VEVs 
$v$, $v_\Phi$ and $v_\De$ in terms of the fundamental parameters in the potential~\cite{Arhrib:2018sbz}.

\subsubsection{Scalar Mass Spectrum \label{subsub5.1.3.2}}
After the electroweak symmetry is broken, we obtained three diagonal blocks in the mass matrix. The first $3\times3$ block with the basis of $S=\{h,\phi_2,\delta_3\}$ is given by
\begin{align}
{\mathcal M}_H^2 =
\begin{pmatrix}
2 \lambda_H v^2 & \lambda_{H\Phi} v v_\Phi 
& \frac{v}{2} \left( M_{H\De} - 2 \lambda_{H \De} v_\De \right)  \\
\lambda_{H\Phi} v v_\Phi
& 
2 \lambda_\Phi v_\Phi^2
&  \frac{ v_\Phi}{2} \left( M_{\Phi\De} - 2 \lambda_{\Phi \De} v_\De \right) \\
\frac{v}{2} \left( M_{H\De} - 2 \lambda_{H \De} v_\De \right)  & \frac{ v_\Phi}{2} \left( M_{\Phi\De} - 2 \lambda_{\Phi \De} v_\De \right) & \frac{1}{4 v_\De} \left( 8 \lambda_\De v_\De^3 + M_{H\Delta} v^2 + M_{\Phi \De} v_\Phi^2 \right)   
\end{pmatrix} \; .
\label{eq:scalarbosonmassmatrix}
\end{align}
This matrix can be diagonalized by an orthogonal matrix $O^{H}$,
\begin{equation}
(O^{H})^{T} \cdot {\cal M}_H^2 \cdot O^H = {\rm Diag}(m_{h_1}^2,m_{h_2}^2,m_{h_3}^2) \; .
\end{equation}
The lightest eigenvalue $m_{h_1}$ is the mass of $h_1$ which is identified as the $125$ GeV Higgs boson observed at the LHC, while $m_{h_2}$ and $m_{h_3}$ are the masses of heavier Higgses $h_2$ and $h_3$ respectively. The physical Higgs $h_i$ ($i =1,2,3$) is 
a mixture of  the three components of $S$: $h_i = O^H_{ji}S_j$. 
Thus the SM-like Higgs boson in this model is a linear combination of the neutral components of the 
two $SU(2)$ doublets $H_1$ and $\Phi_H$ and the real component of the $SU(2)_H$ triplet $\Delta_H$.

The second block is also $3\times3$. In the basis of $G=\{G_H^p, H_2^{0*}, \Delta_p\}$, it is given by 
\begin{align}
{\mathcal M}_D^{ 2} =
\begin{pmatrix}
M_{\Phi \De} v_\De  +\frac{1}{2}\lambda^\prime_{H\Phi}v^2 & \frac{1}{2}\lambda^\prime_{H\Phi}  v v_\Phi & - \frac{1}{2} M_{\Phi \De} v_\Phi  \\
\frac{1}{2}\lambda^\prime_{H\Phi} v v_\Phi &  M_{H \De} v_\De 
+\frac{1}{2}\lambda^\prime_{H\Phi} v_\Phi^2
 &  
\frac{1}{2} M_{H \De} v\\
- \frac{1}{2} M_{\Phi \De} v_\Phi & \frac{1}{2} M_{H \De} v & 
\frac{1}{4 v_\De} \left( M_{H\De} v^2 + M_{\Phi \De} v_\Phi^2 \right)\end{pmatrix} \; .
\label{goldstonemassmatrix}
\end{align}
This matrix can also be diagonalized by an orthogonal matrix $O^{D}$
\begin{equation}
(O^{D})^T \cdot {\cal M}_D^2 \cdot O^D = {\rm Diag} (m_{\tilde{G^p}}^2,m_{D}^2,m_{\De}^2) \; .
\end{equation}
One eigenvalue of Eq.~(\ref{goldstonemassmatrix}) is zero ({\it i.e.} $m_{\tilde{G^p}}=0$) and identified as the unphysical Goldstone boson $\tilde G^p$. The $m_{D}$ and $m_{\tilde{\De}}$ ($m_{D} < m_{\tilde{\De}}$) are masses of two physical fields $D$ and $\tilde{\De}$ respectively. The $D$ could be a DM candidate in G2HDM. 

The final block is a $4 \times 4$ diagonal matrix with the following entries
\begin{align}
m^2_{H^\pm} &=  - \frac{1}{2}\la^\prime_H v^2 +\frac{1}{2}\lambda^\prime_{H\Phi}v_\Phi^2  + M_{H \De} v_\De  \; ,\\
m^2_{G^\pm} &= m^2_{G^0} = m^2_{G^0_H} = 0 \; ,
\label{chargedHiggsmass}
\end{align}
where $m_{H^\pm}$ is mass of the physical charged Higgs $H^\pm$, and 
$m_{G^\pm}, \, m_{G^0}, \, m_{G^0_H}$ are masses of the four Goldstone boson fields $G^\pm$, $G^0$ and $G^0_H$, respectively. 
Note that we have used the minimization conditions Eqs.~\eqref{vevv}, \eqref{vevphi} and \eqref{vevdelta} 
to simplify various matrix elements of the above mass matrices. The six Goldstone particles $G^\pm$, $G^0$, $G^0_H$ and $\widetilde G^{p,m}$ will be absorbed 
by the longitudinal components of the massive gauge bosons 
$W^\pm$, $Z$, $Z^\prime$ and $W^{\prime (p,m)}$ after the electroweak symmetry breaking.
For details of the gauge boson masses, we refer our readers to~\cite{Huang:2015wts,Arhrib:2018sbz}.

\begin{figure}
	   \includegraphics[width=\textwidth]{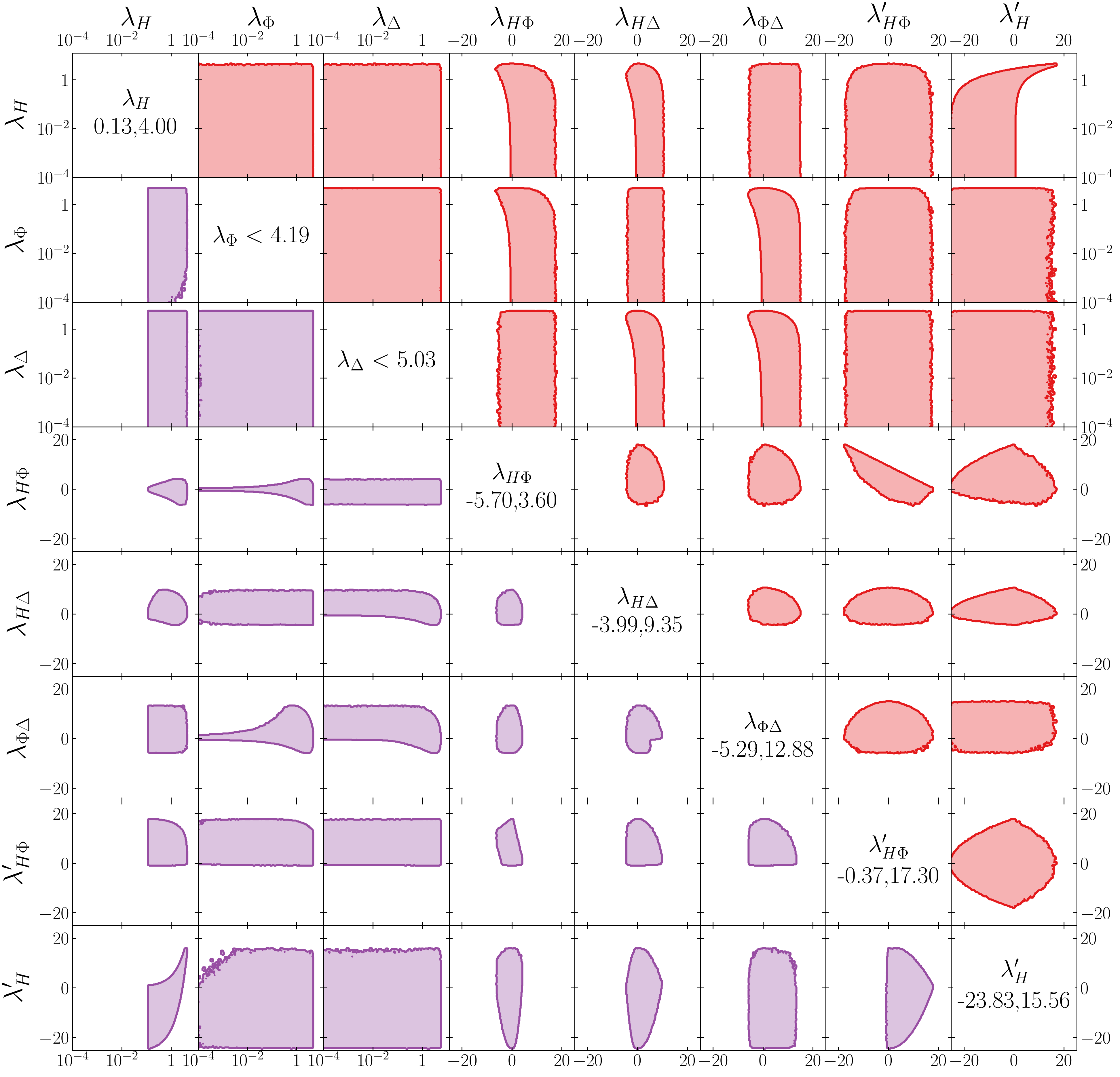}
   \caption{
      \label{fig:matrixplot}
	  A summary of the parameter space allowed by the theoretical and
	  phenomenological constraints. The red regions show the results from
	  the theoretical constraints (VS+PU). The
	  magenta regions are constrained by Higgs physics 
	  as well as the theoretical constraints (HP+VS+PU). Figure taken from \cite{Arhrib:2018sbz}.
  }
\end{figure}

%
\section{Constraints \label{constraint}}
Before presenting our numerical study, we summarize in this section the allowed parameter space 
of the scalar sector of the model. 

The constraints are performed in \cite{Arhrib:2018sbz}, through the requirement of vacuum stability (VS), perturbative unitarity (PU) and Higgs physics (HP), the latter of which includes the Higgs boson mass of 125 GeV 
and signal strengths of Higgs boson decaying into diphoton and $\tau^+\tau^-$ measured at the LHC.
 It should be noted that, in the numerical study of \cite{Arhrib:2018sbz}, the two parameters $M_{H\Delta}$, $M_{\Phi\Delta}$ were set to be varied in the range of $[-1,1]$ TeV, $v_\Delta \in [0.5,20]$ TeV, while $v$ and $v_\Phi$ were fixed to be 246 GeV and 10 TeV, respectively. 
 
We show a summary of allowed regions of parameter space in Fig.~\ref{fig:matrixplot} obtained in \cite{Arhrib:2018sbz}. The diagonal panels indicate the allowed ranges of the eight couplings $\lambda_{H, \Phi, \De}$, $\lambda_{H}^\prime$, and $\lambda_{H\Phi,H\De,\Phi\De}$, $\lambda^\prime_{H\Phi}$ under the combined constraints of (VS+PU+HP). The upper red triangular block
corresponds to (VS+PU) constraints, while the lower magenta triangular block corresponds to the (VS+PU+HP) constraints.
It turns out that among the eight $\lambda-$parameters only two of them $\lambda_H$ and $\lambda_{H\Phi}$ are 
significantly constrained by (VS+PU+HP). 
We note that some of the parameters such as $M_{H\Delta}$, $M_{\Phi\Delta}$ and the VEVs are constrained only by HP but not by (VS+PU).

\section{Higgs Boson Pair Production in G2HDM at the LHC \label{higgspair}}
In the SM, a pair of Higgs bosons can be produced via two channels at the LHC, a triangle loop diagram with Higgs boson as the mediator and a  box loop diagram. However, the small Higgs boson pair production rate, 
which is roughly a thousand times smaller than single Higgs boson
at the 14 TeV LHC \cite{Glover:1987nx,Dawson:1998py,deFlorian:2013uza,deFlorian:2013jea,deFlorian:2015moa, deFlorian:2016uhr,Borowka:2016ypz,Spira:2016ztx,deFlorian:2016spz,Borowka:2016ehy,Plehn:1996wb}, makes the measurement very 
challenging.
We note that a recent combined observed (expected) limit by the ATLAS~\cite{HH95CLLimit} 
on the non-resonant Higgs boson pair cross-section is 0.22 pb (0.35 pb) 
at 95\% confidence level, which corresponds to 6.7 (10.4) times the predicted SM cross-section.

\begin{figure}[hbtp!]
    \centering   
    \begin{subfigure}{0.49\textwidth}
        \includegraphics[width=\textwidth]{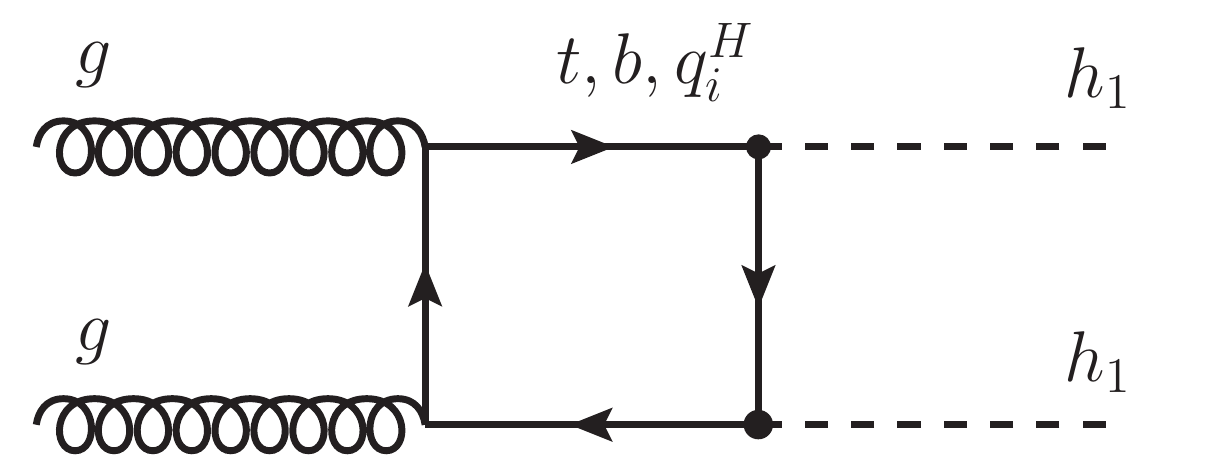}
        \caption{}
        \label{pairHiggsBoxDiag}
    \end{subfigure}
    \begin{subfigure}{0.49\textwidth}
        \includegraphics[width=\textwidth]{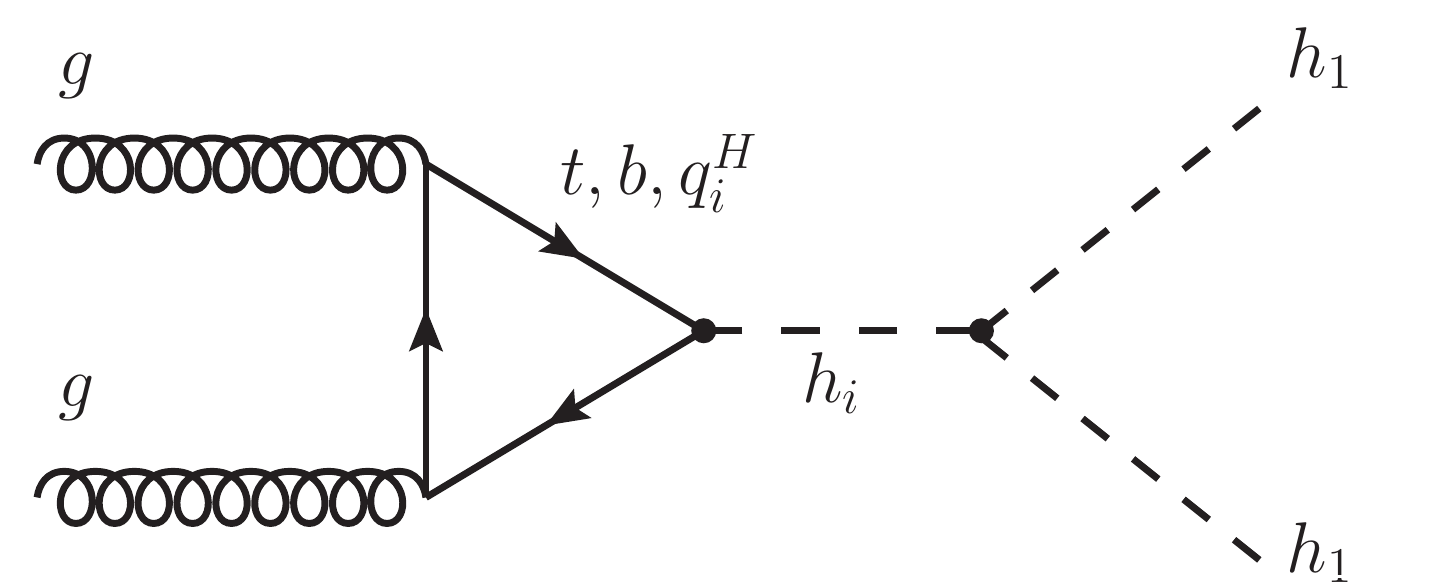}
        \caption{}
        \label{paiHiggsTriagDia}
    \end{subfigure}
 \caption{\small Feynman diagrams for non-resonant (a) and resonant (b) production 
 of a pair of 125 GeV Higgs bosons in G2HDM. Note that $q_i=u,~d,~c,~s,~t,~b$ and $h_i = h_1,~h_2,~h_3$.
 }
 \label{pairHiggsFeyDia}
 \end{figure}

In the G2HDM, the Higgs boson pair production rate can be significantly enhanced since more diagrams contribute, including the new heavy quarks $q_i^H$ in the loop and heavy scalars $h_2, h_3$ as mediators. 
In addition, the 125 GeV SM-like Higgs boson $h_1$ is a mixture of $h$, $\phi_2$ and $\delta_3$, and this mixing has impacts on both modifications in the quark Yukawa couplings and trilinear Higgs self-coupling. Feynman diagrams for production of a pair of $h_1$s in G2HDM are shown in Fig. \ref{pairHiggsFeyDia}.
 The relevant couplings for production of a pair of $h_1$ in G2HDM are listed as follows
 \begin{eqnarray}
 g_{q q h_i} &=& O^H_{1i}\,\frac{m_q}{v} \, , \\
g_{q^H q^H h_i} &=& O^H_{2i}\,\frac{m_{q^H}}{v_{\Phi}}\,, \\
g_{h_1 h_1 h_1} &=& 6 \Big( \lambda_H \,v \,(O^H_{11})^3 + \lambda_{\Phi} \, v_{\Phi} \,(O^H_{21})^3 -  \lambda_{\Delta} \, v_{\Delta} \,(O^H_{31})^3 \Big)\nonumber \\
&+& \frac{3}{2} \Big( \left( M_{H\Delta} - 2 \lambda_{H\Delta}\, v_{\Delta} \right)\,(O^H_{11})^2\,O^H_{31} + \left( M_{\Phi\Delta} - 2 \lambda_{\Phi \Delta}\, v_{\Delta} \right) \,(O^H_{21})^2\,O^H_{31} \Big)  \nonumber \\
&+& 3 (\lambda_{H\Phi}) \,\left( v\,O^H_{11}\,(O^H_{21})^2  + v_{\Phi} (O^H_{11})^2\,O^H_{21} \right) \nonumber \\
&+& 3 \Big( \lambda_{H\Delta} \,v\,O^H_{11}\,(O^H_{31})^2  + \lambda_{\Phi \Delta} \, v_{\Phi} \,O^H_{21} \, (O^H_{31})^2 \Big)  \, , \\
g_{h_2 h_1 h_1} &=& 6 \Big( \lambda_{H} \,v\, (O^H_{11})^2 O^H_{12}  +  \lambda_{\Phi} \,v_{\Phi}\, (O^H_{21})^2 O^H_{22}  -  \lambda_{\Delta} \,v_{\Delta}\, (O^H_{31})^2 O^H_{32} \Big) \nonumber \\
&+& \frac{1}{2} M_{H\Delta} O^H_{11} \big( O^H_{11}  O^H_{32} + 2  O^H_{12}  O^H_{31}\big) + \frac{1}{2} M_{\Phi \Delta} O^H_{21} \big( O^H_{21}  O^H_{32} + 2  O^H_{22}  O^H_{31}\big)  \nonumber \\
&+& \lambda_{H\Delta} \Big[v \bigg( (O^H_{31})^2 O^H_{12} + 2 O^H_{11} O^H_{31} O^H_{32}\bigg) - v_{\Delta} \bigg( (O^H_{11})^2 O^H_{32} + 2 O^H_{11} O^H_{12} O^H_{31}\bigg) \Big] \nonumber \\
&+& \lambda_{\Phi \Delta} \Big[v_{\Phi} \bigg( (O^H_{31})^2 O^H_{22} + 2 O^H_{21} O^H_{31} O^H_{32}\bigg) - v_{\Delta} \bigg( (O^H_{21})^2 O^H_{32} + 2 O^H_{21} O^H_{22} O^H_{31}\bigg) \Big] \nonumber \\
&+&  (\lambda_{H\Phi}) \Big[v \Big((O^H_{21})^2 O^H_{12} + 2 O^H_{11} O^H_{21} O^H_{22}\Big) + v_{\Phi}\Big(O^H_{11} (O^H_{11}O^H_{22} + 2 O^H_{12} O^H_{21})\Big)\Big]  \, , \nonumber\\
\end{eqnarray}
where $g_{qqh_i}$, $g_{q^Hq^Hh_i}$, $g_{h_1 h_1 h_1}$ and $g_{h_2 h_1 h_1}$ are the quark Yukawa couplings, heavy quark Yukawa couplings, trilinear $h_1$ self-coupling and coupling between heavier scalar $h_2$ and two $h_1$s, respectively. One can see that the SM quark Yukawa couplings $g_{qqh_1}$ are now smaller by a factor of the mixing element $O_{11}^H$ as compared to the SM values. Furthermore, the Higgs boson self-couplings $g_{h_1 h_1 h_1}$ and $g_{h_2 h_1 h_1}$ in G2HDM are comprised of many new parameters which might give us a chance to study the effects of these parameters in double $h_1$ production. In what follows, we will ignore the heaviest scalar $h_3$ in our analysis due to its negligible contribution to the double $h_1$ production cross section.

The differential cross section for double $h_1$ production from gluon fusion in G2HDM can be straightforwardly 
translated from the SM formulas~\cite{Plehn:1996wb}, 
\begin{align}
\frac{d \hat{\sigma} (g g \to h_1 h_1)}{d\hat{t}}& = \frac{G_F^2\alpha^2_s}{512\,(2\pi)^3} \nonumber \\
&  \times \Bigg\{ \Big| \sum_{f=q,q^H} \, \sum_{i=1}^{2}\,\kappa_{ffh_i} \, g_{h_ih_1h_1} \, v \,D_{h_i}(\hat{s}) \,F_{\Delta}(\hat{s},\tau_{fi}) + \sum_{f=q,q^H} \,\kappa_{ffh_1}^2\,F_{\Box}(\hat{s},\tau_{fi}) \Big|^2  \nonumber \\
& \;\;\; +  \Big|\, \sum_{f=q,q^H} \, \kappa_{ffh_1}^2\,G_{\Box}(\hat{s},\tau_{fi})\Big|^2 \Bigg\} \; ,
\end{align}
where $G_F$ and $\alpha_s$ are the Fermi constant and strong coupling constant respectively,
\begin{align}
\kappa_{qqh_i} & = O^H_{1i} \, , \\
\label{Cqhqhhi}
\kappa_{q^H q^H h_i} & = O^H_{2i} \frac{v}{v_\Phi} \, , \\
D_{h_i}(\hat{s})&=\frac{1 }{(\hat{s}-m_{h_i}^2+i \,m_{h_i}\Gamma_{h_i})} \,, 
\end{align}
and $F_{\Delta}(\hat{s},\tau_{fi})$, $F_{\Box}(\hat{s},\tau_{fi})$, $G_{\Box}(\hat{s},\tau_{fi})$ with $\tau_{fi} = 4m_f^2/m_{h_i}^2$
are form factors that can be found in the Appendix A.1 of Ref. \cite{Plehn:1996wb}.
For later purpose, we also define $\lambda_{h_ih_1h_1} = g_{h_ih_1h_1}/g^{\rm SM}_{hhh}$ where 
$g^{\rm SM}_{hhh} = 6 \lambda_{\rm SM} v$.
 
\section{Numerical Results \label{NR}}

In this section we will present the numerical analysis for double $h_1$  production in G2HDM at the LHC
and compare the results with SM predictions.

Before we proceed, let us present the set up of the parameter space in the model for scanning.
We will adopt the allowed ranges for all the $\lambda$-parameters, $\lambda_{H, \Phi, \De}$, $\lambda_{H}^\prime$, $\lambda_{H\Phi,H\De,\Phi\De}$, $\lambda^\prime_{H\Phi}$, which satisfy the theoretical constraints from (VS + PU) obtained in \cite{Arhrib:2018sbz}.
Recall that these theoretical constraints are only relevant for the quartic couplings in the scalar potential.
For the Higgs phenomenology constraints, other parameters in the scalar potential are also involved,
and we will vary their ranges as follows 
\begin{eqnarray}
0.1 \,{\rm GeV} \,<& \,v_{\Delta} < & \, 4\,\mathrm{TeV} \; ,\\
30 \,{\rm TeV} \,<& \,v_{\Phi} \,< &\, 100 \,{\rm TeV} \; , \\
-3 \,{\rm TeV} \,<& \,M_{H \Delta} &\,<\, 3\,\mathrm{TeV} \; ,\\
0 \,< & \,M_{\Phi\Delta} & <\, 15\,\mathrm{GeV} \; .
\end{eqnarray}
The SM VEV $v$ is fixed at 246 GeV.
\begin{figure}[hbtp!]
    \centering   
    \begin{subfigure}{0.49\textwidth}
        \includegraphics[width=\textwidth]{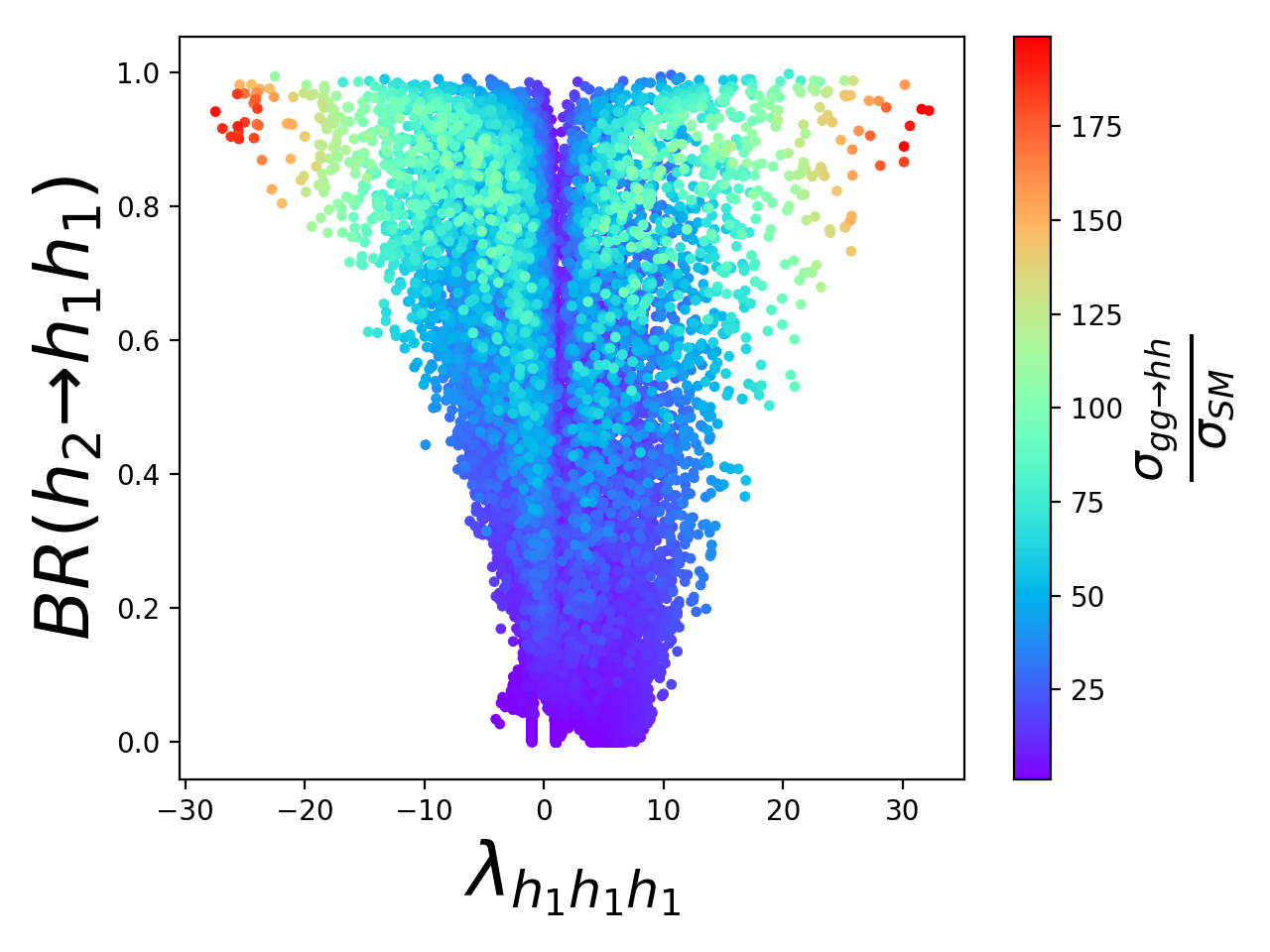}
        \caption{}
        \label{pairHiggs_woDM_1}
    \end{subfigure}
    \begin{subfigure}{0.49\textwidth}
        \includegraphics[width=\textwidth]{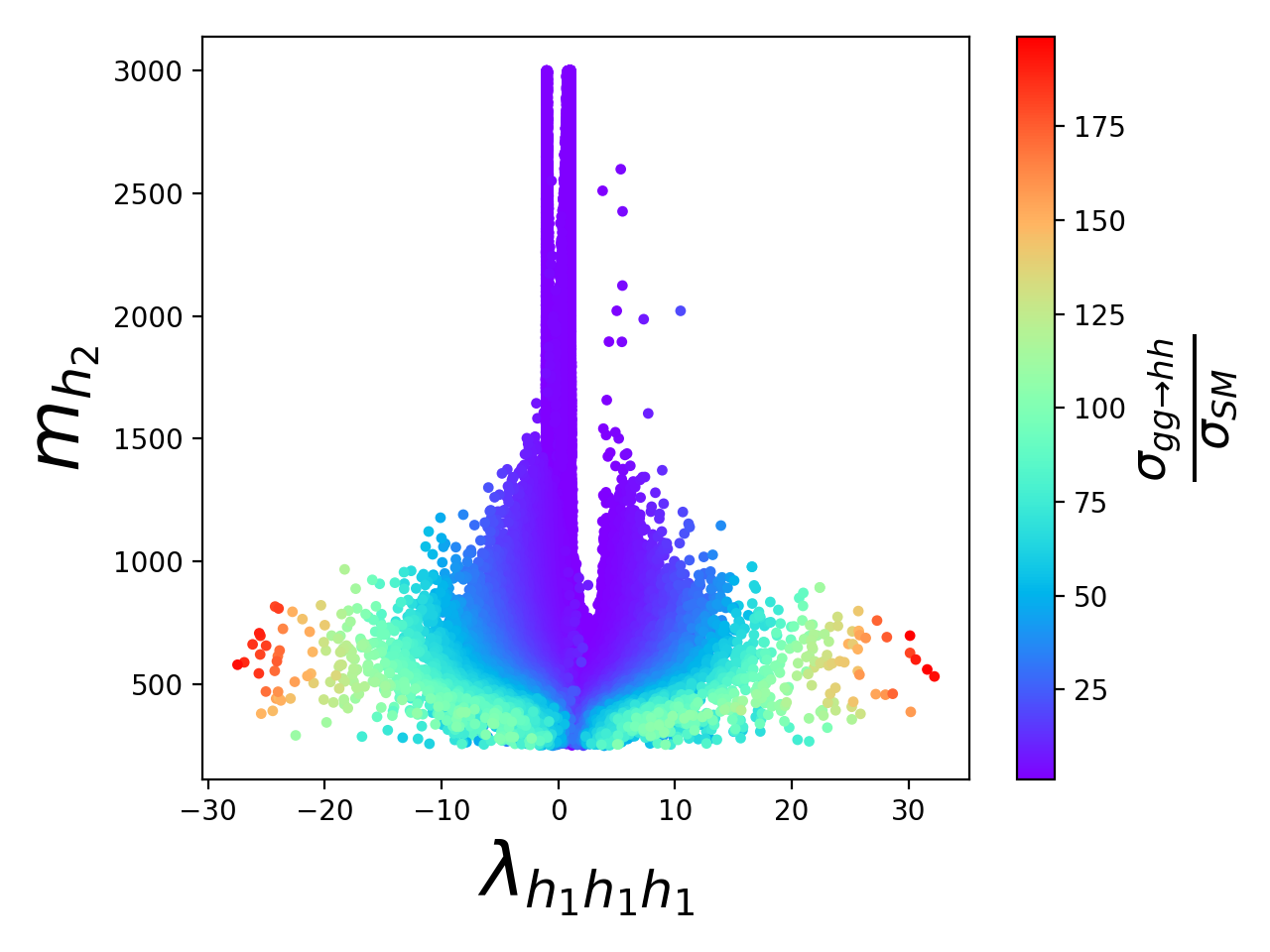}
        \caption{}
        \label{pairHiggs_woDM_2}      
     \end{subfigure}  
     \begin{subfigure}{0.49\textwidth}
        \includegraphics[width=\textwidth]{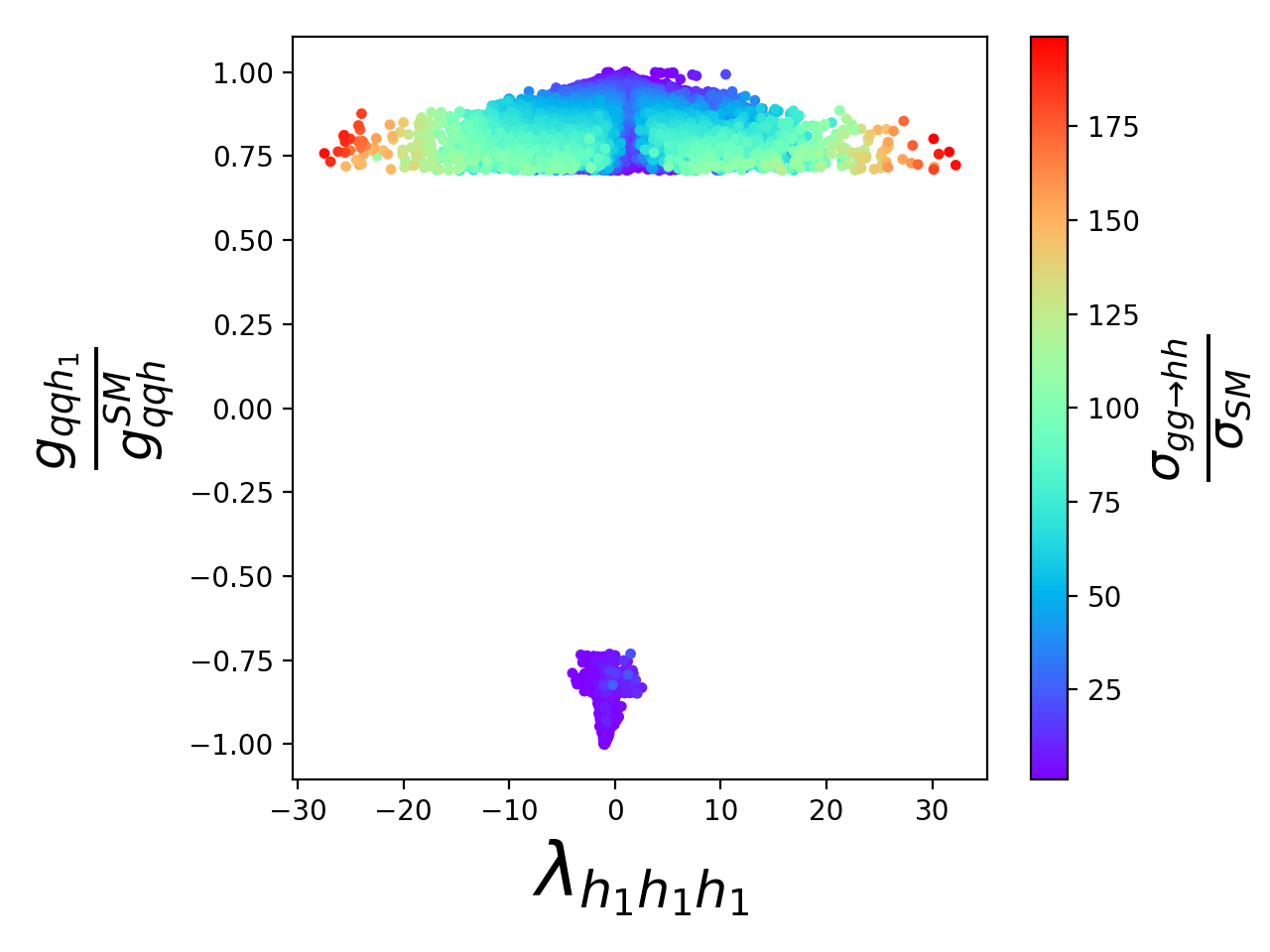}
        \caption{}
        \label{pairHiggs_woDM_3}
     \end{subfigure}
      \begin{subfigure}{0.49\textwidth}
        \includegraphics[width=\textwidth]{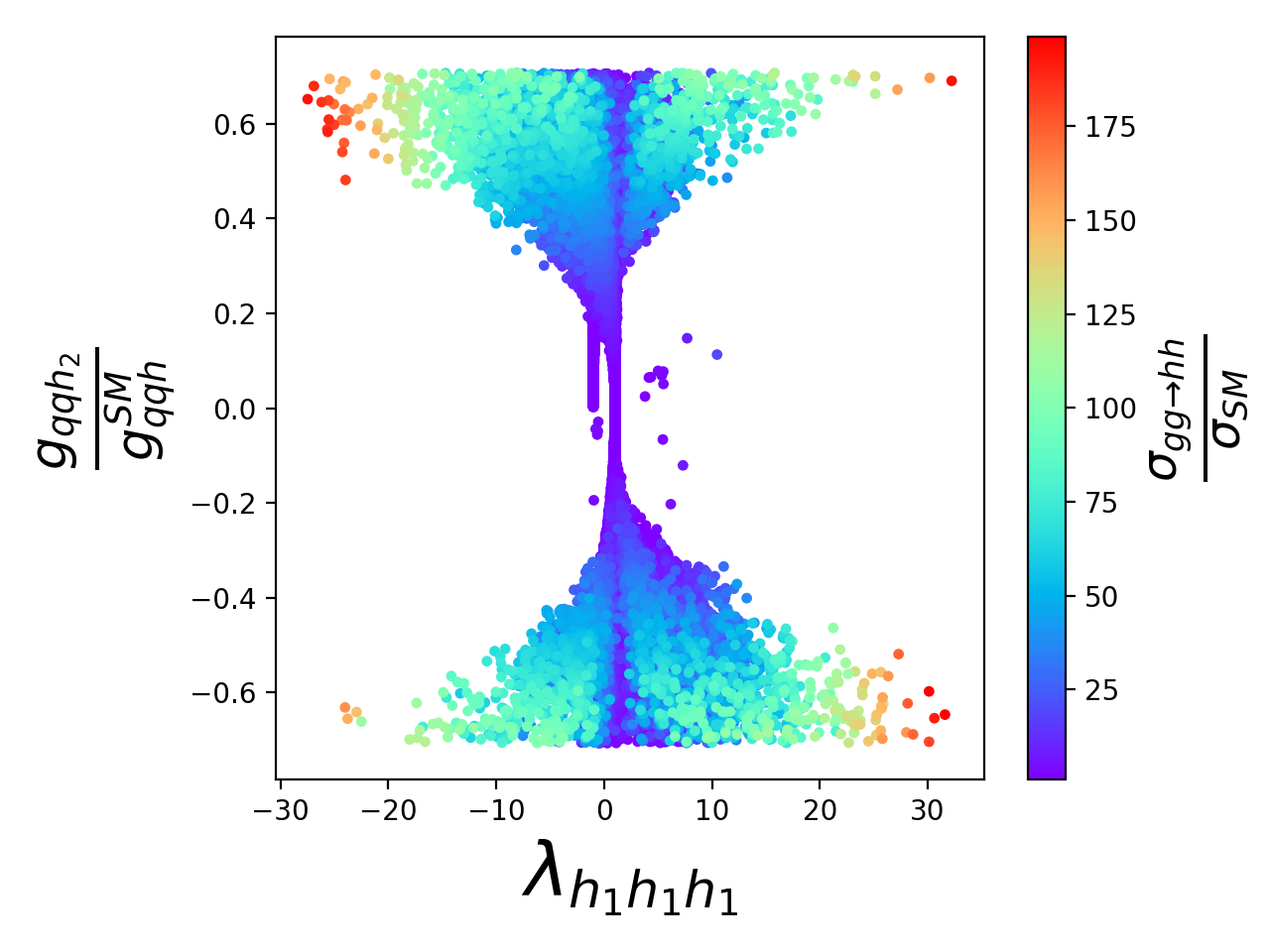}
        \caption{}
        \label{pairHiggs_woDM_4}        
    \end{subfigure}
 \caption{\small The scatter plots of relevant parameters to Higgs boson pair production without the experimental 
 constraints from DM relic density and direct searches. The color palette 
 indicates the ratio of double Higgs boson production cross sections between G2HDM and SM.
 Note that  $\lambda_{h_1 h_1 h_1} = g_{h_1 h_1 h_1} / g_{hhh}^{\rm SM}$ with 
 $g_{hhh}^{\rm SM} = 6 \lambda_{\rm SM} v$, $\kappa_{qqh_1} = q_{qqh_1}/q_{qqh}^{\rm SM}$ and 
  $\kappa_{qqh_2} = q_{qqh_2}/q_{qqh}^{\rm SM}$ with $q_{qqh}^{\rm SM} = m_q / v$.
  }
 \label{pairHiggs_woDM}
 \end{figure}
First, we scan all the parameters with the set-up ranges defined above and require them to 
pass through all the theoretical and Higgs phenomenological 
constraints presented in \cite{Arhrib:2018sbz}. 
The constraints from direct $Z^\prime$ resonance search at the latest ATLAS and CMS 13 TeV results \cite{Aaboud:2017yvp, Aaboud:2016cth, CMS:2016abv, CMS:2016wpz} have been taken into account in our scanning. Furthermore, the dark matter candidate $D$ is set to be heavier than half of the Higgs boson mass so that the invisible mode of $h_1 \to DD$ is not kinematically allowed.
The masses of heavy fermions are assumed to be degenerate and set to be 3 TeV. Finally, we focus on the situation $m_{h_2} > 2 m_{h_1}$ to allow $h_2$ decays 
on shell into $h_1 h_1$.

In Fig.~\ref{pairHiggs_woDM}, we show the scatter plots of the ratio of production cross sections 
for a pair of 125 GeV Higgs bosons
between the G2HDM and SM on the planes of  
($\lambda_{h_1h_1h_1}$, ${\rm BR}(h_2 \to h_1 h_1)$) (Fig. \ref{pairHiggs_woDM_1}), 
($\lambda_{h_1h_1h_1}$, $m_{h_2}$) (Fig. \ref{pairHiggs_woDM_2}), 
($\lambda_{h_1h_1h_1}$, $\kappa_{qqh_1}$) (Fig. \ref{pairHiggs_woDM_3})
and ($\lambda_{h_1h_1h_1}$, $\kappa_{qqh_2}$) (Fig.  \ref{pairHiggs_woDM_4}).
The color palette on the right of each of the plots  
in Fig.~\ref{pairHiggs_woDM} indicates the signal strength of the double $h_1$ Higgs boson production.
From these four plots in Fig. \ref{pairHiggs_woDM}, one can see that the trilinear self-coupling of Higgs boson in G2HDM can significantly deviate from SM value, it can even flips its sign to be negative. 
From Fig. \ref{pairHiggs_woDM_1}, one observes that
the branching ratio of the heavier scalar $h_2$ decay into a pair of $h_1$s can vary from 0 up to $100\%$. As expected, when $|\lambda_{h_1h_1h_1}|$ and ${\rm BR}(h_2 \to h_1 h_1)$ are getting larger, the triangle diagram will become the dominant channel and enhance the production cross section. Note that when $\lambda_{h_1h_1h_1}$ becomes negative, there is a constructive interference between the two types of box and triangle Feynman diagrams in Fig. \ref{pairHiggsFeyDia}. However when one of the channels, either the box or triangle Feynman diagram, becomes the dominant contribution to the total production cross section, the interference effect is not significant anymore. It is also shown in Fig. \ref{pairHiggs_woDM_2} that for a heavier $h_2$ mass 
the cross section of double $h_1$ production will be much smaller.
Furthermore, due to the constraints from the Higgs physics, the absolute value of Yukawa couplings of SM quarks with $h_1$ can not be deviated too much from its SM value which is demonstrated in Fig. \ref{pairHiggs_woDM_3}, while Fig. \ref{pairHiggs_woDM_4} shows the Yukawa couplings between SM quark and Higgs boson $h_2$ could be small due to the smallness of mixing between $SU(2)_L$ doublet scalar $H$ and  $SU(2)_H$ doublet scalar $\Phi_H$. 
The contributions from the heavy quarks in G2HDM are found to be small because the Yukawa couplings 
$\kappa_{q^Hq^Hh_i}$ in~\eqref{Cqhqhhi} are scaled by the small VEV ratio $v/v_\Phi$.
 
\begin{figure}[hbtp!]
    \centering   
    \begin{subfigure}{0.49\textwidth}
        \includegraphics[width=\textwidth]{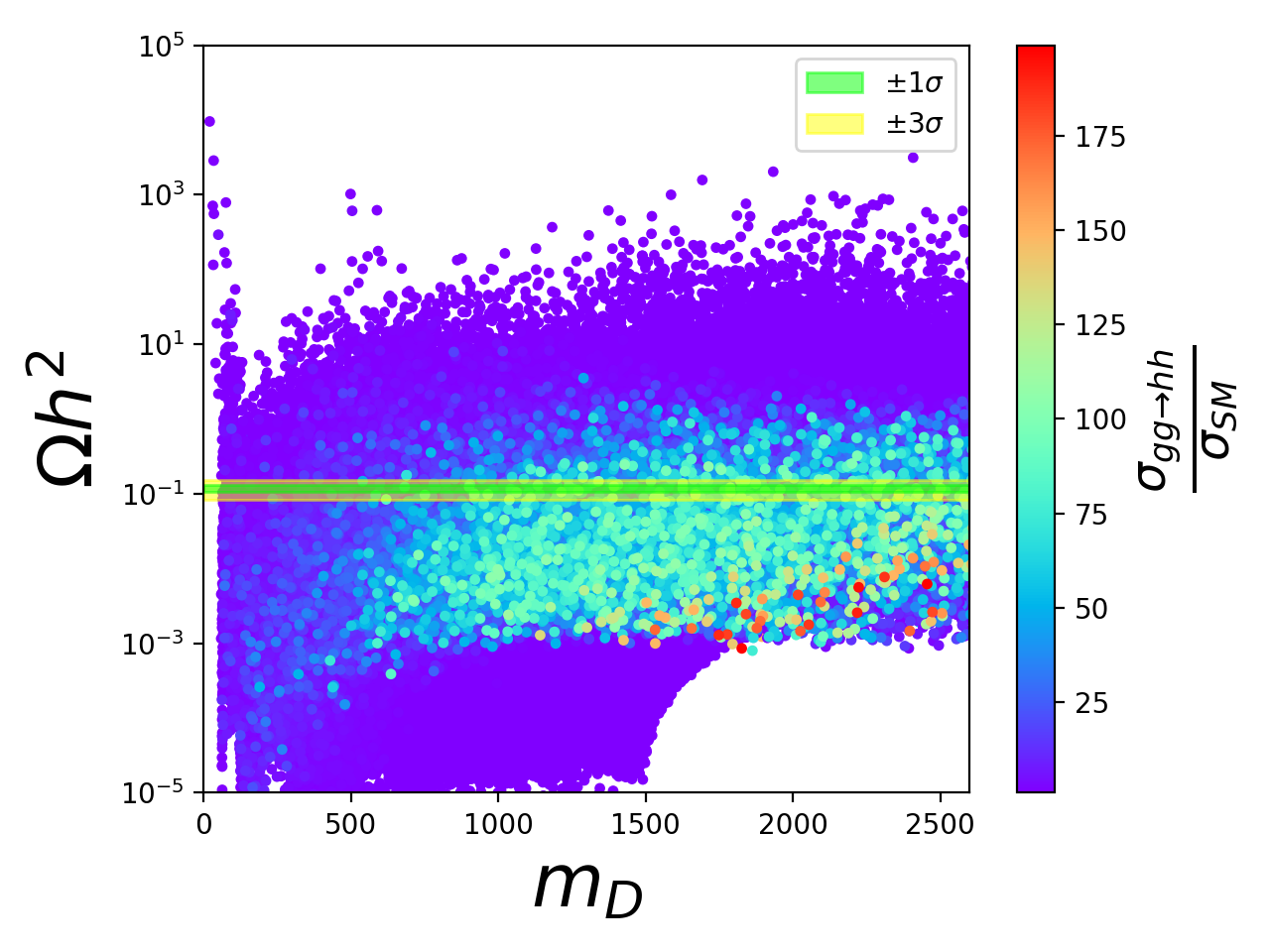}
        \caption{}
        \label{pairHiggs_DM_5}
    \end{subfigure}
    \begin{subfigure}{0.49\textwidth}
        \includegraphics[width=\textwidth]{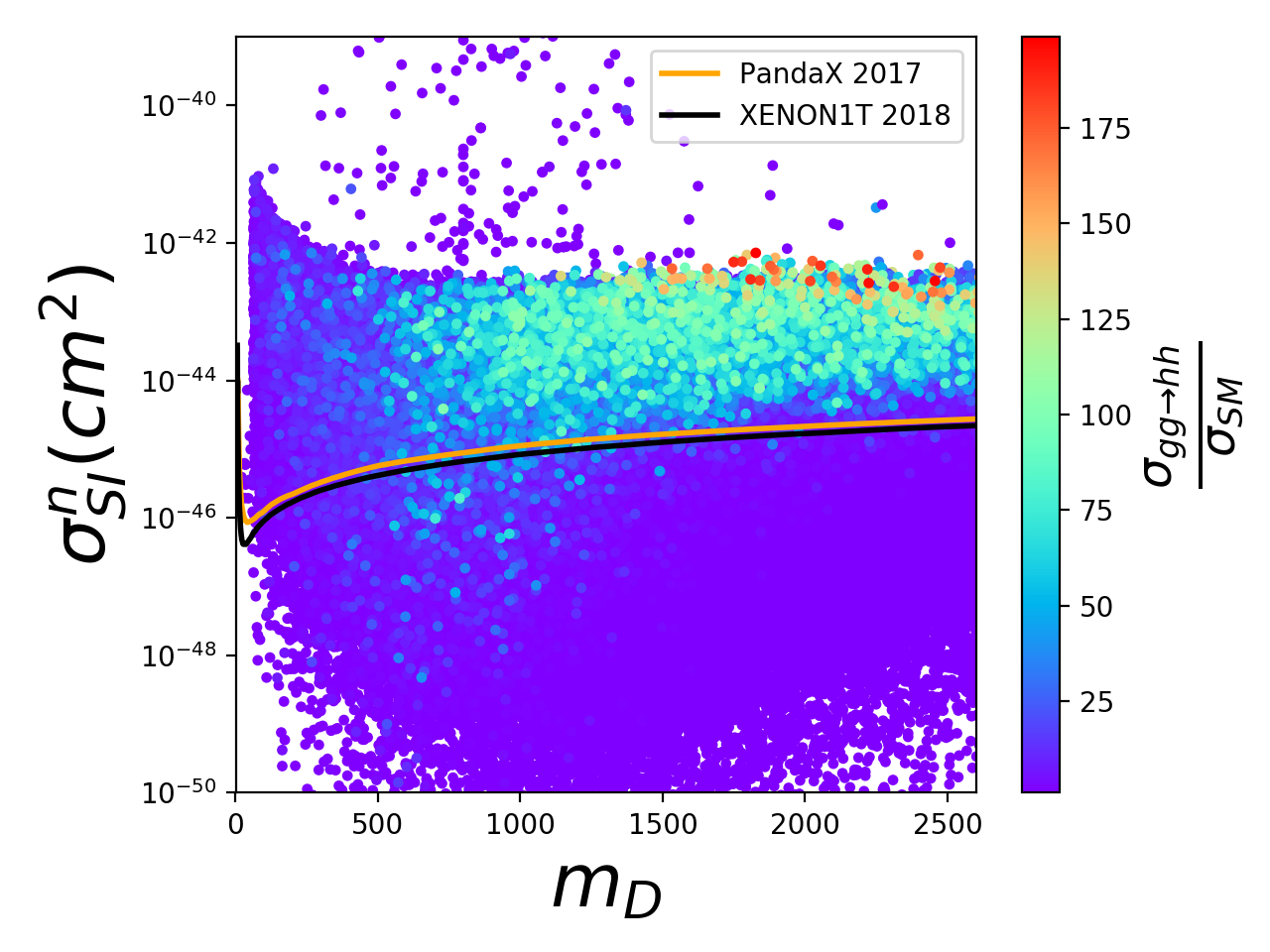}
        \caption{}
        \label{pairHiggs_DM_6}      
     \end{subfigure}  
  \caption{\small The scatter plots for the ratio of production cross sections for a pair of 125 GeV Higgs bosons between G2HDM and SM on the planes of dark matter mass and a) relic density of DM, b) spin-independent cross section of DM and nucleon. The lime (yellow) band corresponds to $1\sigma$ ($3\sigma$) range of the PLANCK's relic density measurement of DM \cite{Ade:2015xua}. The orange and black lines represent the upper limit on spin-independent cross section of DM and nucleon from PandaX-II Experiment~\cite{Cui:2017nnn} and XENON1T~\cite{Aprile:2018dbl}, respectively.
  }
 \label{pairHiggs_DM}
 \end{figure}

Since dark matter candidate exists in G2HDM, we consider further the dark matter constraints from the cosmological observations and direct search experiments. 
We used the \texttt{MadDM} package \cite{Backovic:2015cra} to calculate the relic density of the DM candidate and its elastic scattering cross sections with nucleon. 
In Fig. {\ref{pairHiggs_DM}}, we present the scatter plots for the ratio of production cross sections for a pair of 125 GeV Higgs bosons between G2HDM and SM on plane of the dark matter mass and a) relic density of DM, b) spin-independent cross section of DM and nucleon. The lime (yellow) band corresponds to $1\sigma$ ($3\sigma$) range of the PLANCK's relic density measurement of DM \cite{Ade:2015xua}. The orange and black line represent the upper limit on spin-independent cross section of DM and nucleon from PandaX-II Experiment~\cite{Cui:2017nnn} and XENON1T~\cite{Aprile:2018dbl}, respectively. Imposing the mass of dark matter candidate $D$ to be the lightest among $\tilde{\De}$, $W^{\prime (p,m)}$, $H^{\pm}$ and heavy fermions implies $m_D$ to be less than $\sim 2.7$ TeV. In the region of $m_D > 500$ GeV, there are correlations between Higgs boson pair production cross section and DM relic density as well as DM-nucleon cross section. In particular, the cross section of gluon-gluon fusion to double $h_1$ tends to be larger when DM relic density becomes smaller or DM-nucleon cross section becomes larger. 
The first correlation, shown in Fig. \ref{pairHiggs_DM_5}, is due to the fact that the $|\lambda_{h_1h_1h_1}|$ and ${\rm BR}(h_2 \to h_1 h_1)$ can control not only the Higgs boson pair production but also DM annihilation cross section. Indeed, when they both become bigger, the DM annihilation process will be dominated by $D D \to h_i \to h_1h_1$ channel, implying the DM annihilation cross section becomes larger or DM relic density becomes smaller. 
The second correlation, shown in Fig. \ref{pairHiggs_DM_6}, is due to the fact that the DM-nucleon cross section has about half its contributions coming from the one loop heavy quarks (mainly top quark) in the triangle diagram which also appear in the double Higgs boson production process.

The DM relic density and direct searches put stringent constraints on the parameter space of G2HDM. 
As shown in Fig. {\ref{pairHiggs_DM_5}}, the PLANCK's relic density measurement constrains 
the parameter space in a small $3\sigma$ band, while from Fig. {\ref{pairHiggs_DM_6}}
one can also see that the DM direct search constraints cut off almost 
all the parameter space which significantly enhances the cross section of double Higgs boson production.
Moreover, when both relic density and direct search constraints are imposed, only about $2\%$ of the data points 
survived. 

  \begin{figure}
    \centering   
    \begin{subfigure}{0.49\textwidth}
        \includegraphics[width=\textwidth]{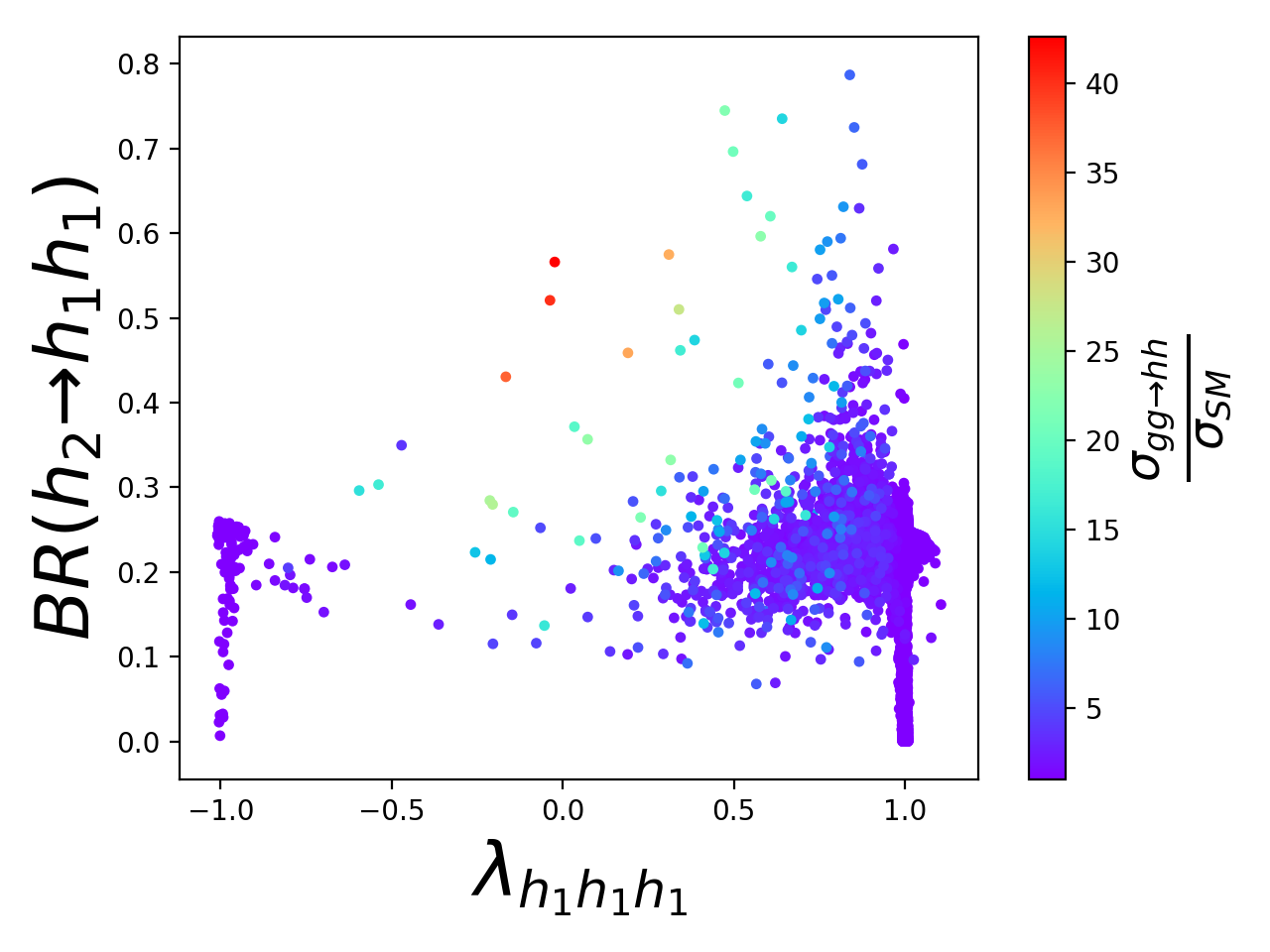}
        \caption{}
        \label{pairHiggs_wDM_1}
    \end{subfigure}
    \begin{subfigure}{0.49\textwidth}
        \includegraphics[width=\textwidth]{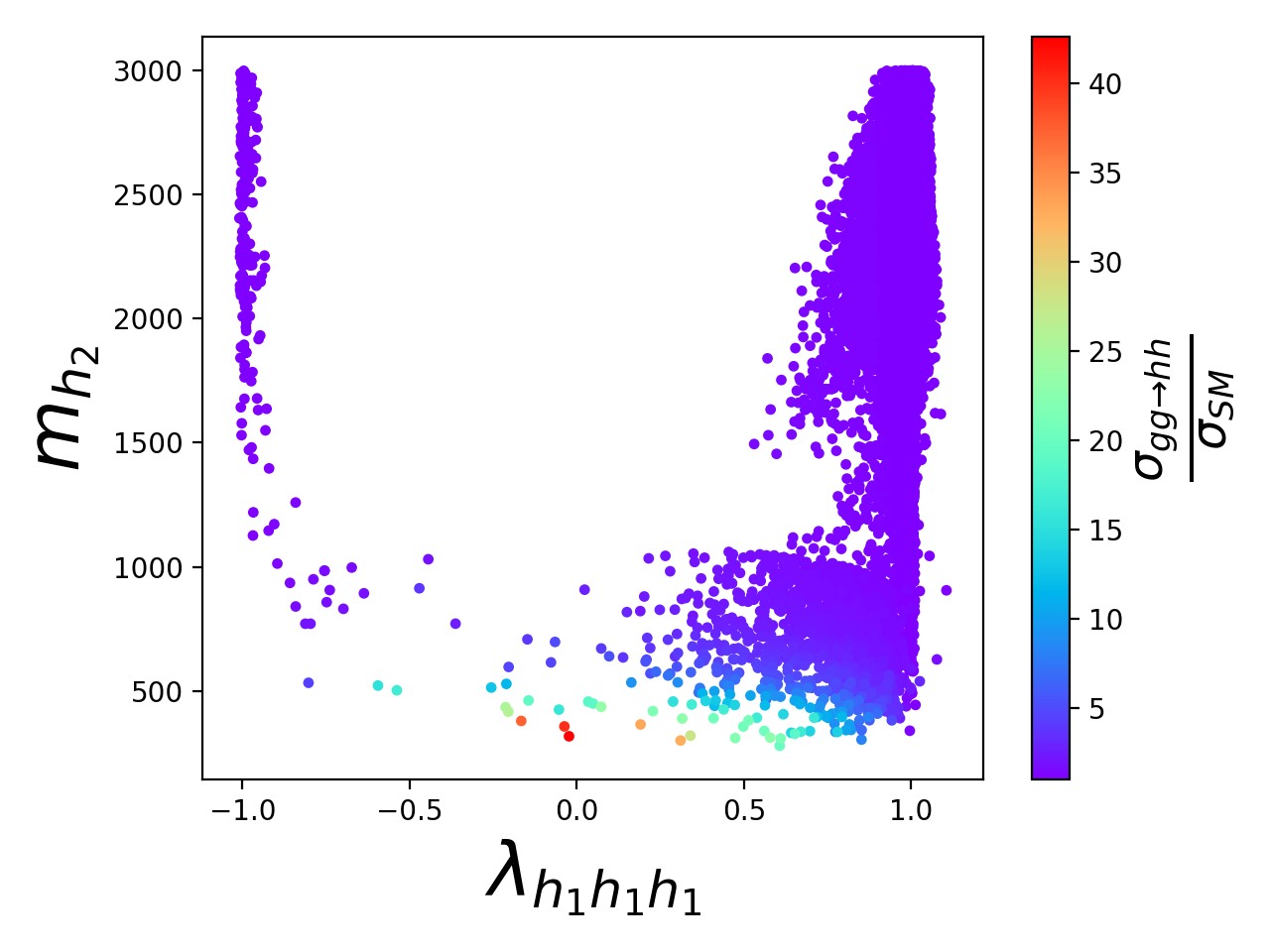}
        \caption{}
        \label{pairHiggs_wDM_2}      
     \end{subfigure}  
     \begin{subfigure}{0.49\textwidth}
        \includegraphics[width=\textwidth]{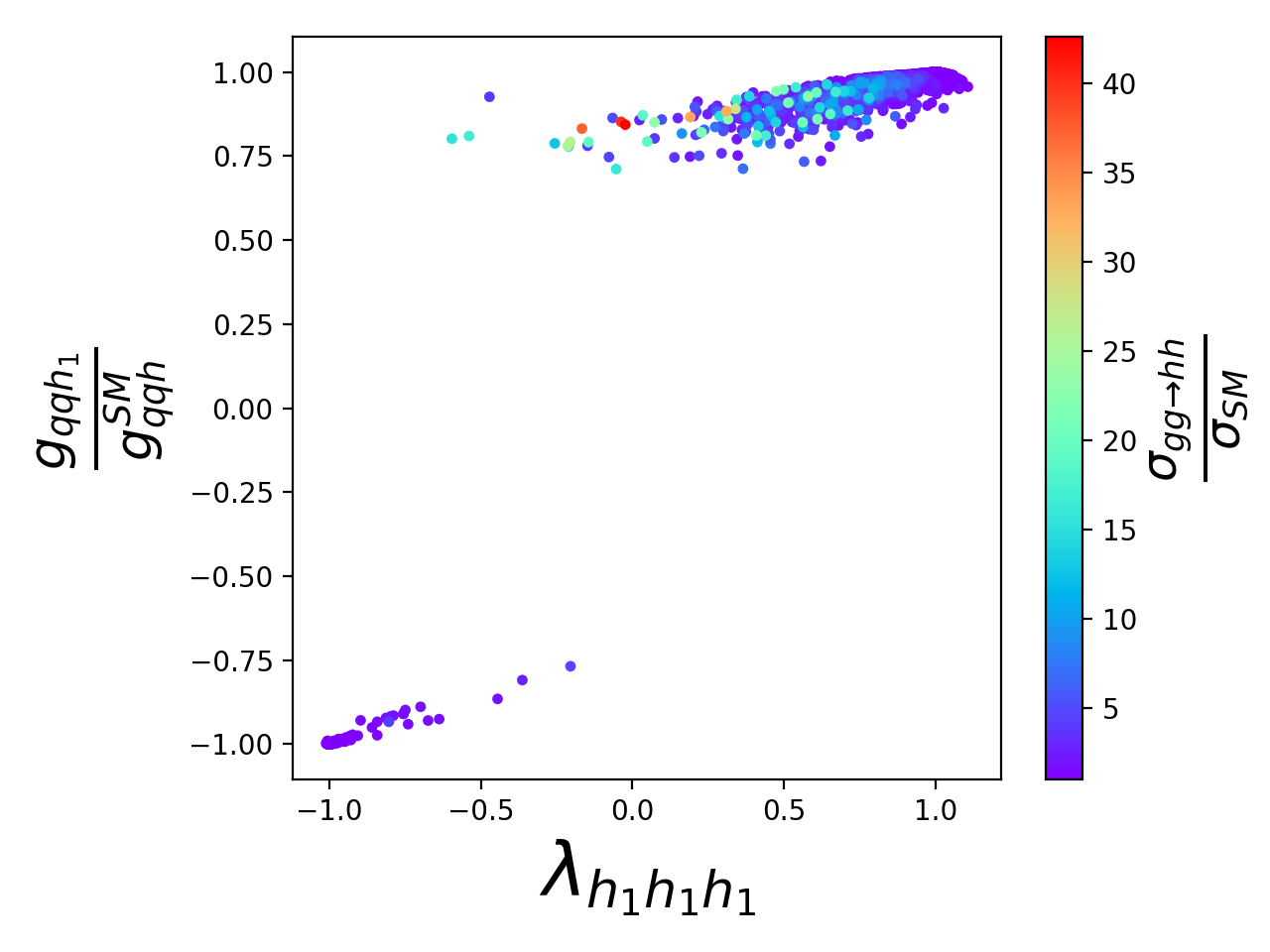}
        \caption{}
        \label{pairHiggs_wDM_3}
     \end{subfigure}
      \begin{subfigure}{0.49\textwidth}
        \includegraphics[width=\textwidth]{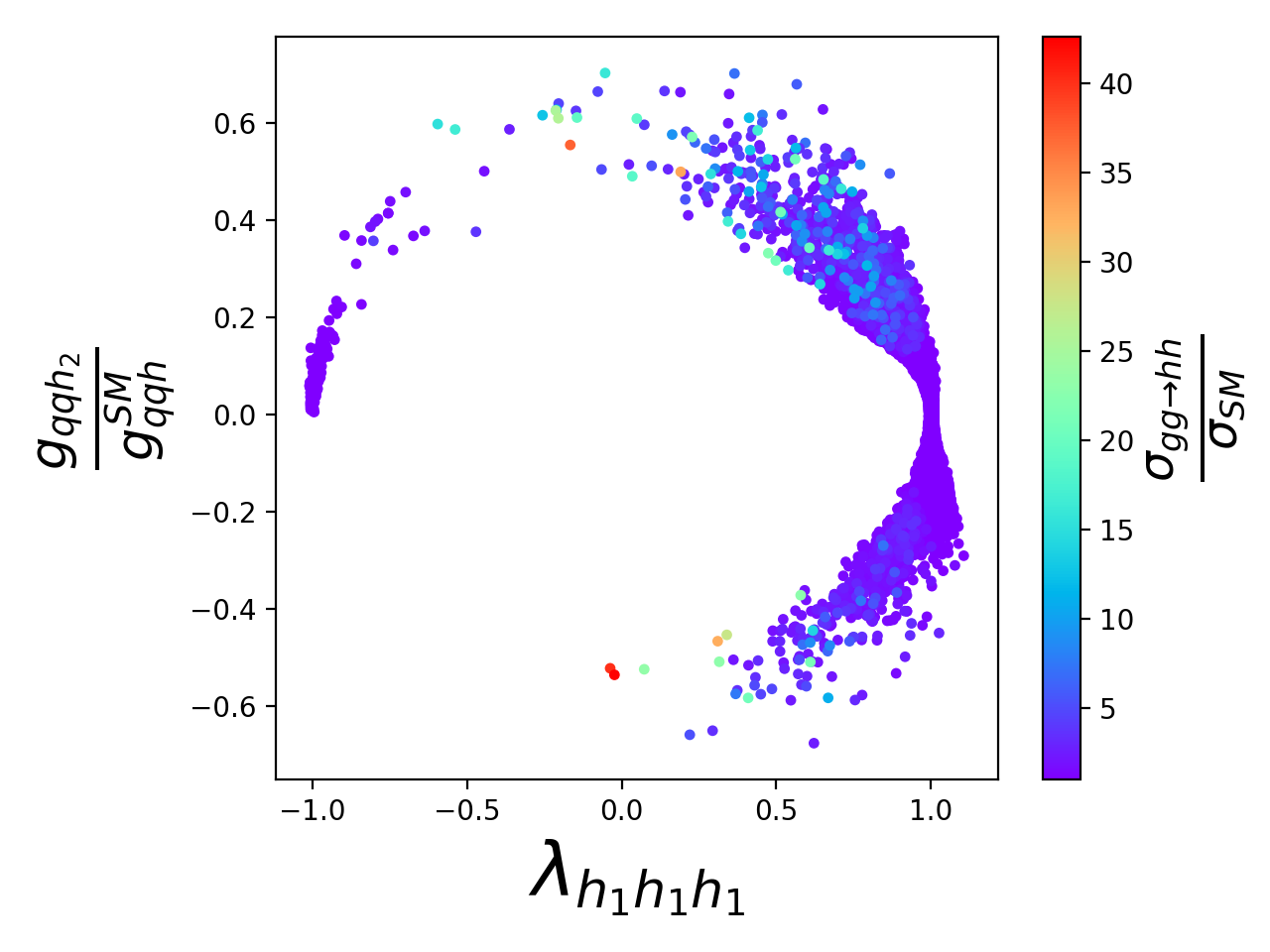}
        \caption{}
        \label{pairHiggs_wDM_4}        
    \end{subfigure}
 \caption{\small Same as Fig. {\ref{pairHiggs_woDM}} but after taking into account the experimental 
 constraints from DM relic density from PLANCK~\cite{Ade:2015xua} 
 and direct searches from PandaX-II Experiment~\cite{Cui:2017nnn} and XENON1T~\cite{Aprile:2018dbl}.
 }
 \label{pairHiggs_wDM}
 \end{figure}
    
Same as Fig.~{\ref{pairHiggs_woDM}}, we show in Fig.~\ref{pairHiggs_wDM} the scatter plots of 
relevant parameters to Higgs boson pair production 
after taking into account the constraints from DM relic density and direct searches.
The allowed points in the parameter space are selected within $3 \sigma$ of the PLANCK's relic density measurement of DM \cite{Ade:2015xua} and below the upper limits of the DM direct detection searches from 
PandaX-II experiments~\cite{Cui:2017nnn} and XENON1T~\cite{Aprile:2018dbl}. 
Under these combined constraints of (VS+PU+HP+DM), the parameter space can be narrowed down further. 
For example, we now have the $-1\leq \lambda_{h_1h_1h_1} \leq 1.3$ and ${\rm BR}(h_2 \to h_1 h_1)$ is less than about $80\%$. The negative value of $\lambda_{h_1h_1h_1}$ gives an enhancement of the production cross section because the constructive interference occurs between box and triangle diagrams. Overall, the production cross section of $h_1$ pair is about one order of magnitude lower as compared 
with the one before imposing the DM constraints.  

To do further analysis, 
we pick seven benchmark points 
from the final allowed parameter space satisfying the (VS+PU+HP+DM) constraints, 
at which the mass of heavier scalar $h_{2}$ varying from 300 to 900 GeV. 
In Table~\ref{BP}, we show the fundamental parameters in the scalar potential, derived couplings, 
mass spectra of the scalars, 
and the signal strength for Higgs boson pair production at each benchmark point. 
For benchmark point A, the trilinear self-coupling $\lambda_{h_1h_1h_1}$ is about the same as SM value. 
For the points C and D, we have the negative values for $\lambda_{h_1h_1h_1}$ which can lead to constructive interference 
between the box and triangle diagrams, 
while point B is chosen in which large branching ratio of $h_2 \to h_1 h_1$ can be achieved to see the enhancement effects 
of heavy scalar resonance on the production cross section (See Table~\ref{BRh2}). 
For the points E, F and G, the production cross section is about twice of its SM value. 
One can see that the Yukawa couplings between SM quarks and the $h_1$ Higgs boson are close to SM values in all benchmark points except benchmark point D. 
\begin{table}
\caption{Seven benchmark points allowed by the combined (VS+PU+HP+DM) constraints.}
\centering
\resizebox{\textwidth}{!}
{\begin{tabular}{|c|c|c|c|c|c|c|c|}
\hline
Benchmark point & A & B & C & D & E & F & G \\
\hline
\hline
$\lambda_H$ &  0.35& 0.60 & 0.80&1.79  &0.66  & 2.49 &2.22 \\
$\lambda_\Phi$ & 2.75 & 1.83 & 1.43& 2.45 &  1.44&  1.68&  3.43\\
$\lambda_\Delta$ & 0.84 & 0.37 &2.43 & 0.05 & 0.67 & 1.97 & 0.08\\
$\lambda_H^\prime$ & $-$3.78 &  $-$0.75& $-$6.55 & $-$2.52 & $-$17.80 &1.31  &0.45 \\
$\lambda_{H \Phi}$ & $-$1.37 & 1.41& $-$0.05 & $-$2.24 &  0.003& 0.83& $-$2.15\\
$\lambda_{H \Delta}$ & $-$0.75 & 1.30 & 0.034& $-$0.53 &  $-$0.316& 0.86 & $-$0.31 \\
$\lambda_{\Phi \Delta}$ & 3.06 & 2.11 & 3.78& 0.73 & 2.08 & 3.64 &0.95 \\
$\lambda_{H \Phi}^\prime$ &6.04  & 6.94&7.59 & 7.41& 1.46 & 0.40& 6.16\\
$v_\Delta \, ({\rm GeV})$ & 1926 &1793& 3378 & 621& 1520 & 3212&3458\\
$v_\Phi \, ({\rm GeV})$ & 36220 & 36274 &41580& 30800 & 51914 & 86229 & 33348\\
$M_{H\Delta}\, ({\rm GeV})$ & 199.6 & 2203 &  1625 &  1117& $-$2293 & 2214& $-$2986 \\
$M_{\Phi \Delta}\, ({\rm GeV})$ & 0.91 & 8.72 &11.09 & 0.50 & 1.80 & 0.64 & 3.56\\
\hline
$\lambda_{h_1h_1h_1}$ & 0.85 & 0.15& $-0.53$ & $-0.20$ & 0.84 &  0.35 & 0.41 \\
$\lambda_{h_2h_1h_1}$ & 0.76 & 3.03 & 3.88 & 3.25 & $-3.32$ & 5.42 & $-7.16$  \\
$\kappa_{qqh_1}$          & 0.95 & 0.91 & $0.81$ & $-0.77$ & 0.93& 0.75& 0.86\\
$\kappa_{qqh_2}$          &0.29  & 0.41 & 0.58 & 0.64& $-0.37$ &0.65& $-0.52$ \\
$\kappa_{q^Hq^Hh_1}$ & $-5\times10^{-5}$&  $-10^{-4}$& $3.7\times10^{-4}$ & $-10^{-5}$ & $4\times10^{-5}$& $-8\times10^{-5}$& $7\times10^{-5}$\\
$\kappa_{q^Hq^Hh_2}$ & $2\times10^{-4}$&  $1.7\times10^{-4}$& $5.1\times10^{-4}$ & $4\times10^{-5}$ & $9\times10^{-5}$& $9\times10^{-5}$& $8\times10^{-5}$\\
\hline
$m_{h_2} ({\rm GeV})$ & 300 & 400 & 500 & 600 & 700 & 800 & 900\\
$m_{h_3} ({\rm TeV})$ & 85 & 69.49 & 70.77& 68.22 & 88.35 & 158.2 & 87.39\\
$m_{D} ({\rm GeV})$ & 398 & 1278 & 1210 & 467 & 883 & 619 & 553\\
$m_{\tilde \Delta} ({\rm TeV})$ & 62.94 & 67.61 &81.03 & 59.29 & 44.38 & 38.87 & 58.45\\
$m_{H^\pm} ({\rm TeV})$ & 62.94 & 67.60 & 81.03& 59.29 & 44.39 & 38.87 &58.44 \\
\hline
\hline
$\frac{\sigma(gg\to h_1h_1)}{\sigma_{\rm SM}}$ & 8.2  & 27.3 & 16.7 & 4.6 & 2.1 & 2.1& 2.1\\
\hline
\end{tabular}}
\label{BP}
\end{table}
In Table~\ref{BRh2}, we show the branching ratio of $h_2$ decays to all two body final states in our benchmark points. We observe that the heavy scalar $h_2$ mainly decays into a pair of SM-like Higgs boson $h_1$, $W$ and $Z$ bosons, and top quark. 
\begin{table}
\caption{Branching ratios of the two body decays of $h_2$ in the seven benchmark points.}
\centering
{\begin{tabular}{|c|c|c|c|c|c|c|c|c|}
\hline
Benchmark point & A & B & C& D &E&F&G\\
\hline
 $h_2 \to h_1 h_1$ & 0.329    & 0.575   & 0.298   & 0.113   & 0.175  & 0.100  & 0.161\\
$h_2 \to W^+ W^-$ & 0.462    & 0.255   & 0.391   & 0.496   &0.471   & 0.529 & 0.500\\
$h_2 \to Z Z$          & 0.206    & 0.119   & 0.186   & 0.240   & 0.230  & 0.260 & 0.247 \\
$h_2 \to t \bar{t} $ & 0           & 0.049   & 0.123   & 0.150   & 0.122  & 0.114 & 0.091\\
$h_2 \to b \bar{b}$ &$\sim0$ & $\sim0$ &$\sim0$& $\sim0$ & $\sim0$ & $\sim0$ & $\sim0$ \\
\hline
\end{tabular}}
\label{BRh2}
\end{table}

In order to perform detailed simulations at the LHC, we first implement the G2HDM model into the \texttt{FeynRules} package \cite{Alloul:2013bka} and pass the UFO model files into \texttt{MadGraph5} aMC@NLO \cite{Alwall:2014hca} to generate the events of $h_1$ pair production. 
The Higgs boson decay is done with \texttt{MadSpin} \cite{Artoisenet:2012st} package, and we focus on two decay modes, $h_1 \to b\bar{b}$ and $h_1 \to \gamma \gamma$. 
Finally, the \texttt{Pythia8} package \cite{Sjostrand:2007gs} is used for parton showering and hadronization the events, while the \texttt{Delphes3} package \cite{deFavereau:2013fsa} (with ATLAS setting) 
is used as the fast detector simulation. 

In the next two subsections, we will concentrate on the $b\bar{b}\gamma \gamma$and  $b\bar{b} b\bar{b}$ final states. According to the current searches of the Higgs boson pair production at LHC, the $b\bar{b}\gamma \gamma$ final state channel is a good channel search for the lower mass regime of the heavy scalar~\cite{TheATLAScollaboration:2016ibb, CMS:2017ihs}, while the  $b\bar{b} b\bar{b}$ search channel has better sensitivity for the search for the heavier mass regime of the heavy scalar~\cite{ATLAS:2016ixk, CMS:2016foy}. Thus we use our benchmark points A, B, C, D to study the $b\bar{b}\gamma \gamma$ final state channel, while E, F, G are used for $b\bar{b} b\bar{b}$ final state channel.

\begin{figure}[hbtp!]
    \centering   
    \begin{subfigure}{0.49\textwidth}
        \includegraphics[width=\textwidth]{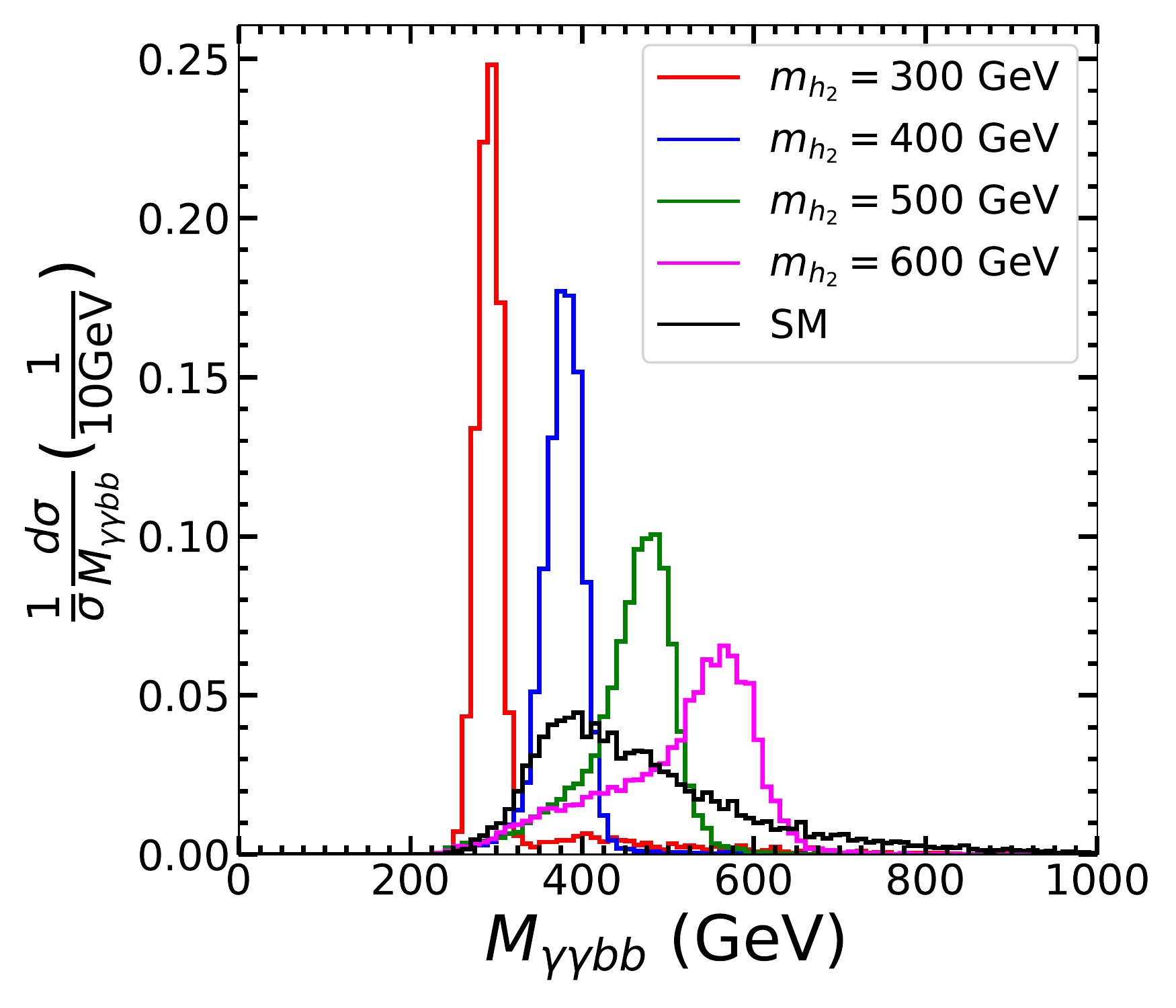}
        \caption{}
        \label{Maabb}
    \end{subfigure}
    \begin{subfigure}{0.49\textwidth}
        \includegraphics[width=\textwidth]{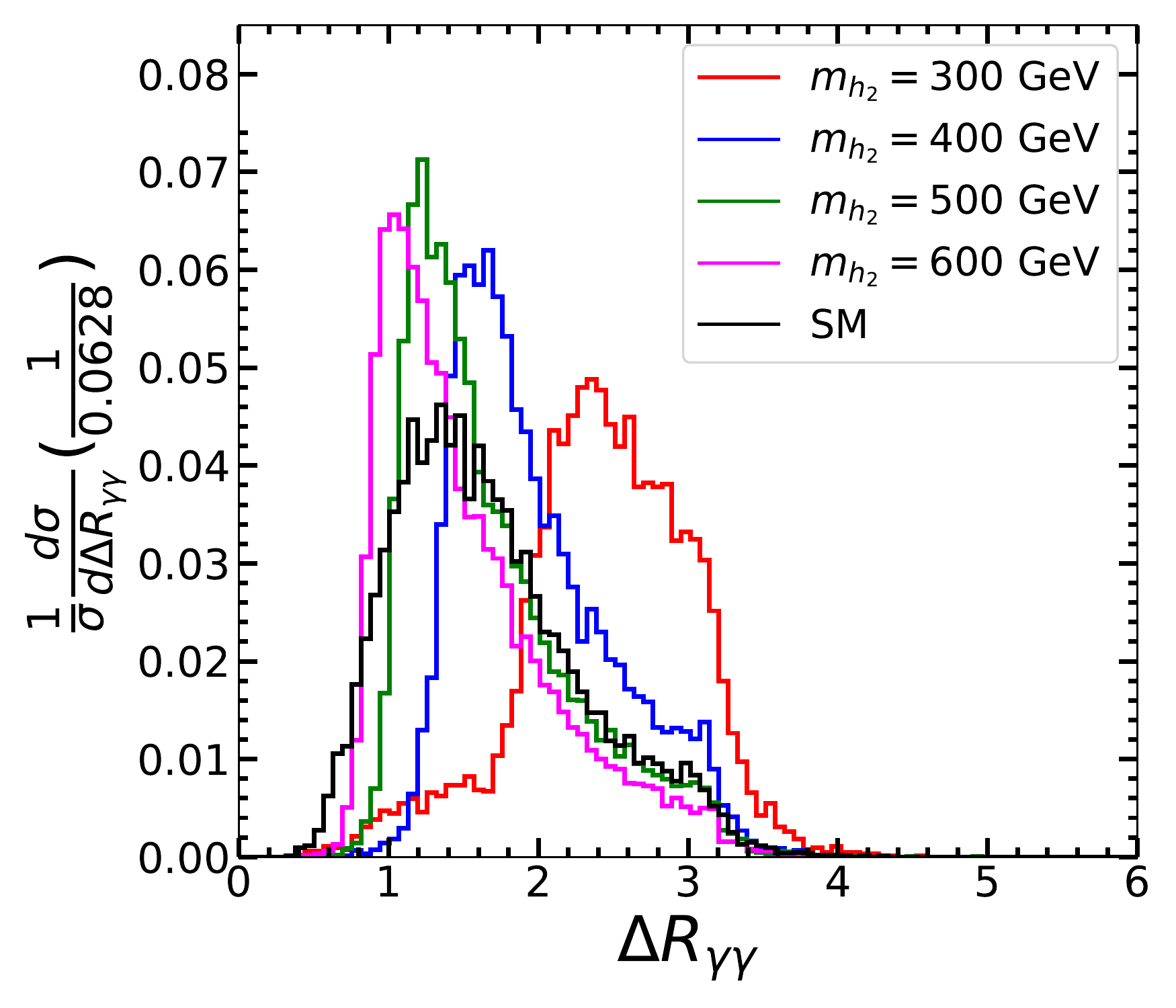}
        \caption{}
        \label{dRaa_aabb}
    \end{subfigure}
    \begin{subfigure}{0.49\textwidth}
        \includegraphics[width=\textwidth]{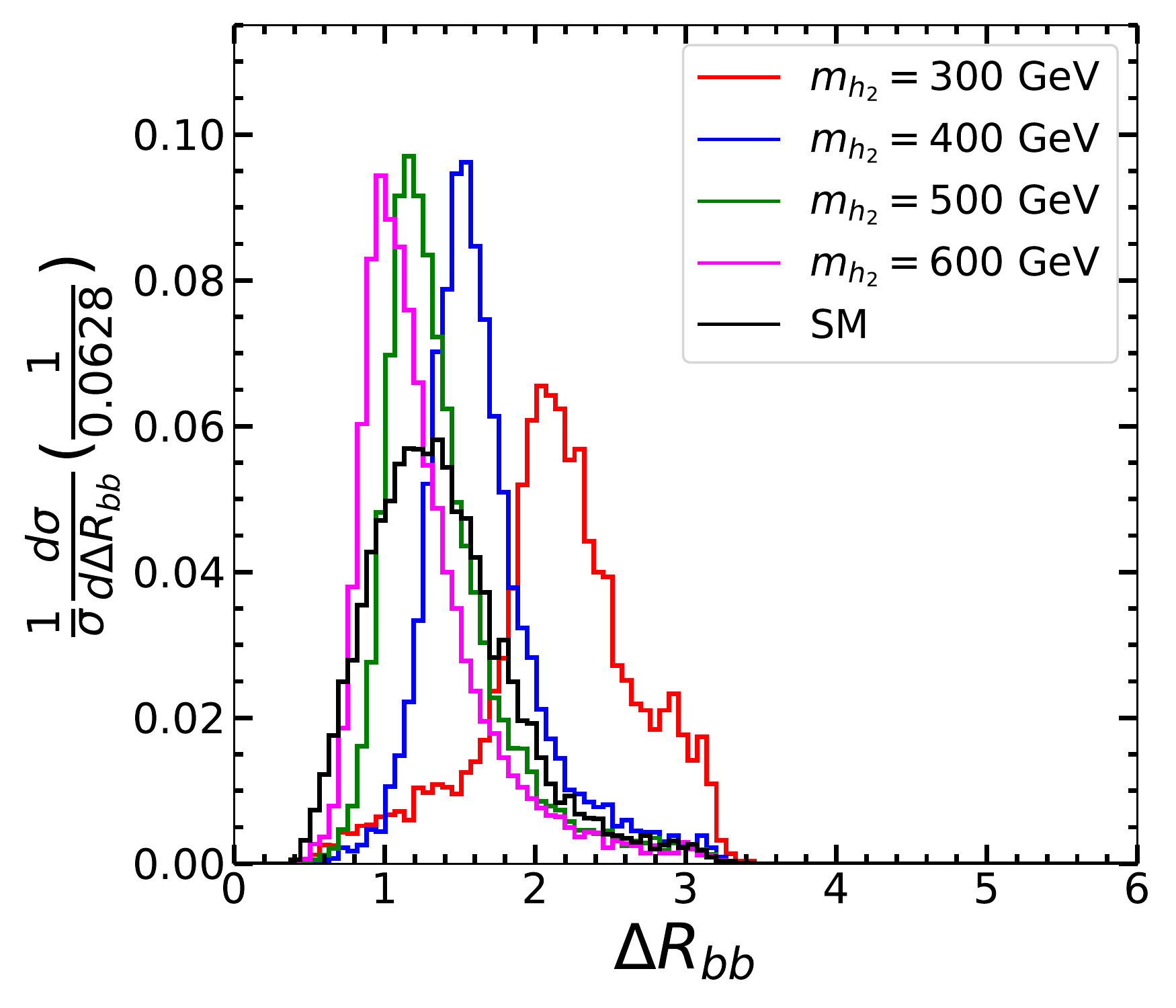}
        \caption{}
        \label{dRbb_aabb}
     \end{subfigure}
     \caption{\small The kinematic distributions of the $b\bar{b}\gamma\gamma$ channel for a) the invariant mass 
     $M_{\gamma\gamma bb}$ of $b\bar{b}\gamma\gamma$, b) the opening angle $\Delta R_{\gamma\gamma}$ of two photons 
     and c) the opening angle $\Delta R_{bb}$ of 2 $b$-jets in SM and in G2HDM 
     with $m_{h_2} = $ 300, 400, 500 and 600 GeV at $\sqrt{s} = 13 \,\rm{TeV}$ LHC.}
     \label{pairHiggs_aabb}
\end{figure}

\subsection{The $b\bar{b}\gamma\gamma$ Final State Channel}
\label{subsub5.2.3.1}

In this section, we perform the kinematic distributions of the $b\bar{b}\gamma\gamma$ final state. We select 4 benchmark points A, B, C, D, at which the mass of heavy scalar $h_2$ equals 300, 400, 500, 600 GeV respectively, to study this channel. Here, we follow the cuts used in the ATLAS experiment for $b\bar{b}\gamma\gamma$ channel analysis \cite{TheATLAScollaboration:2016ibb}, which we summarize as follows:
\begin{itemize}
\item
First, we isolate photons with opening angle ${\Delta}R = 0.2$.
\item
Next, using the anti-$k_T$ algorithm jets are reconstructed with cone radius $R = 0.4$ 
and required to have $\left| \eta \right| < 2.5$ and $p_T > 25$ GeV.
\item
We require at least 2 photons and exactly 2 jets with $b$-tagging efficiency set 
at the default value of \texttt{MadGraph5}.
\item
The leading $b$-jet must have  transverse momentum $p_T > 55 \,\rm{GeV}$, while the sub-leading $b$-jet is required to have $p_T > 35 \,\rm{GeV}$.
\item
Furthermore, the diphoton invariant mass $m_{\gamma\gamma}$ is required to lie between 105 GeV and 160 GeV and the $b$-jet pair invariant mass $m_{b\bar{b}}$ is required to fall into a mass window of 95 GeV to 135 GeV. 
\end{itemize}

In Fig.~{\ref{pairHiggs_aabb}}, we present the kinematic distributions of the $b\bar{b}\gamma\gamma$ channel for a) the invariant mass of $b\bar{b}\gamma\gamma$, b) the opening angle $\Delta R_{\gamma\gamma}$ 
of two photons and c) the opening angle $\Delta R_{bb}$ 
of 2 $b$-jets in SM and in G2HDM with $m_{h_2} = $ 300, 400, 500 and 600 GeV 
at the LHC with $\sqrt{s} = 13 \,\rm{TeV}$. In Fig.~{\ref{Maabb}}, it is obvious to see that the invariant mass distributions peaked at the corresponding mass of heavy scalar $h_2$, while the peak around 400 GeV is for the SM. The peaks are getting lower when the mass of $h_2$ becomes heavier, this is due to the fact that when the mass of $h_2$ becomes heavier the non-resonant process is getting more relevant to the total production cross section. One can also observe that the opening angles of the two photons and of the pair of $b$-jets in Fig.~{\ref{dRaa_aabb}} and Fig.~{\ref{dRbb_aabb}} respectively, tend to be narrower when the mass of $h_2$ becomes heavier. This is 
expected because when the parent decaying particle $h_2$ becomes heavier, the two daughter $h_1$ Higgs bosons 
will be more boosted, implying that the opening angles $\Delta R_{\gamma\gamma}$ and $\Delta R_{bb}$ would be smaller.  

 \begin{figure}
    \centering   
    \begin{subfigure}{0.49\textwidth}
        \includegraphics[width=\textwidth]{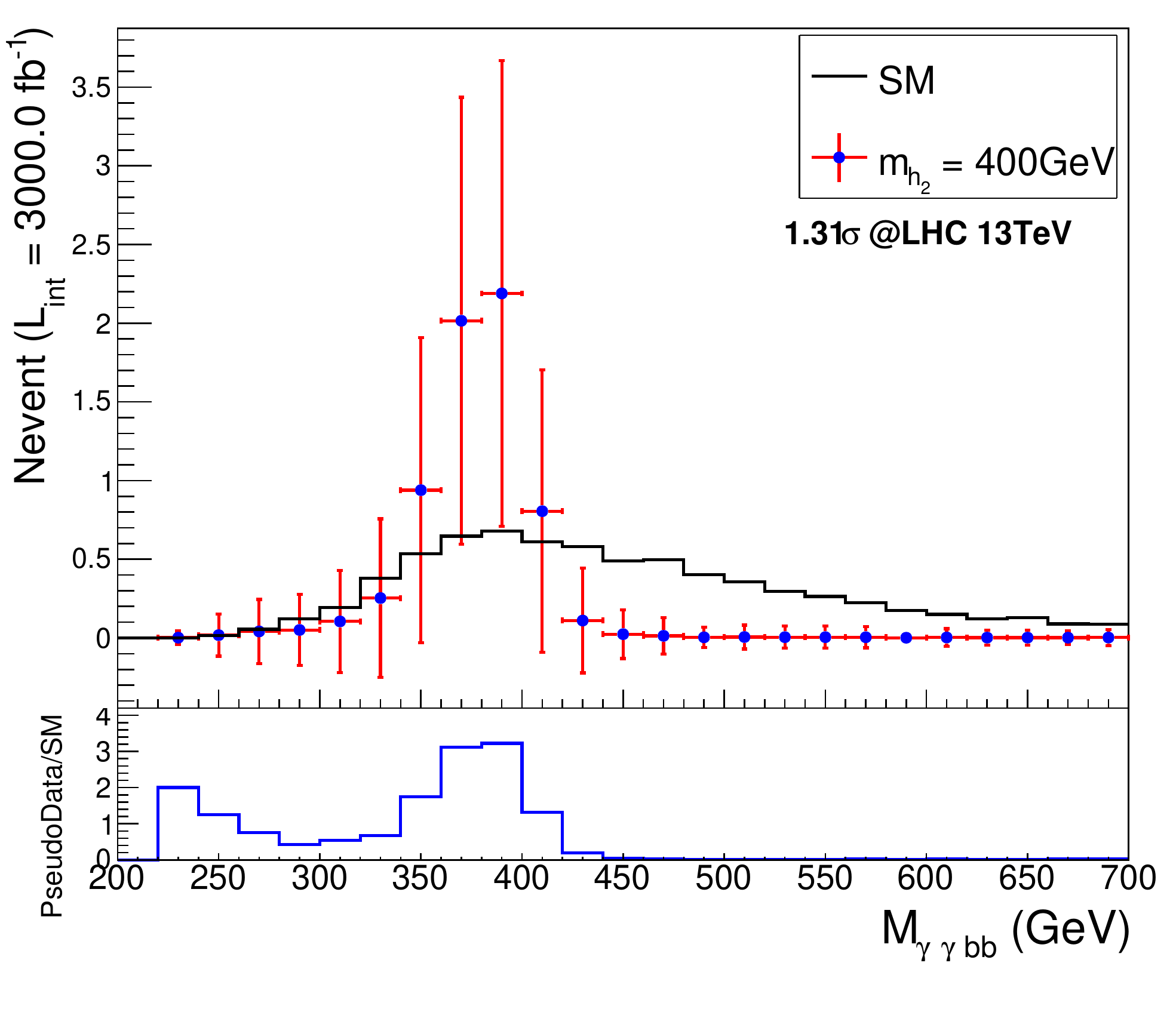}
        \caption{}
        \label{Maabb_L3k}
    \end{subfigure}
    \begin{subfigure}{0.49\textwidth}
        \includegraphics[width=\textwidth]{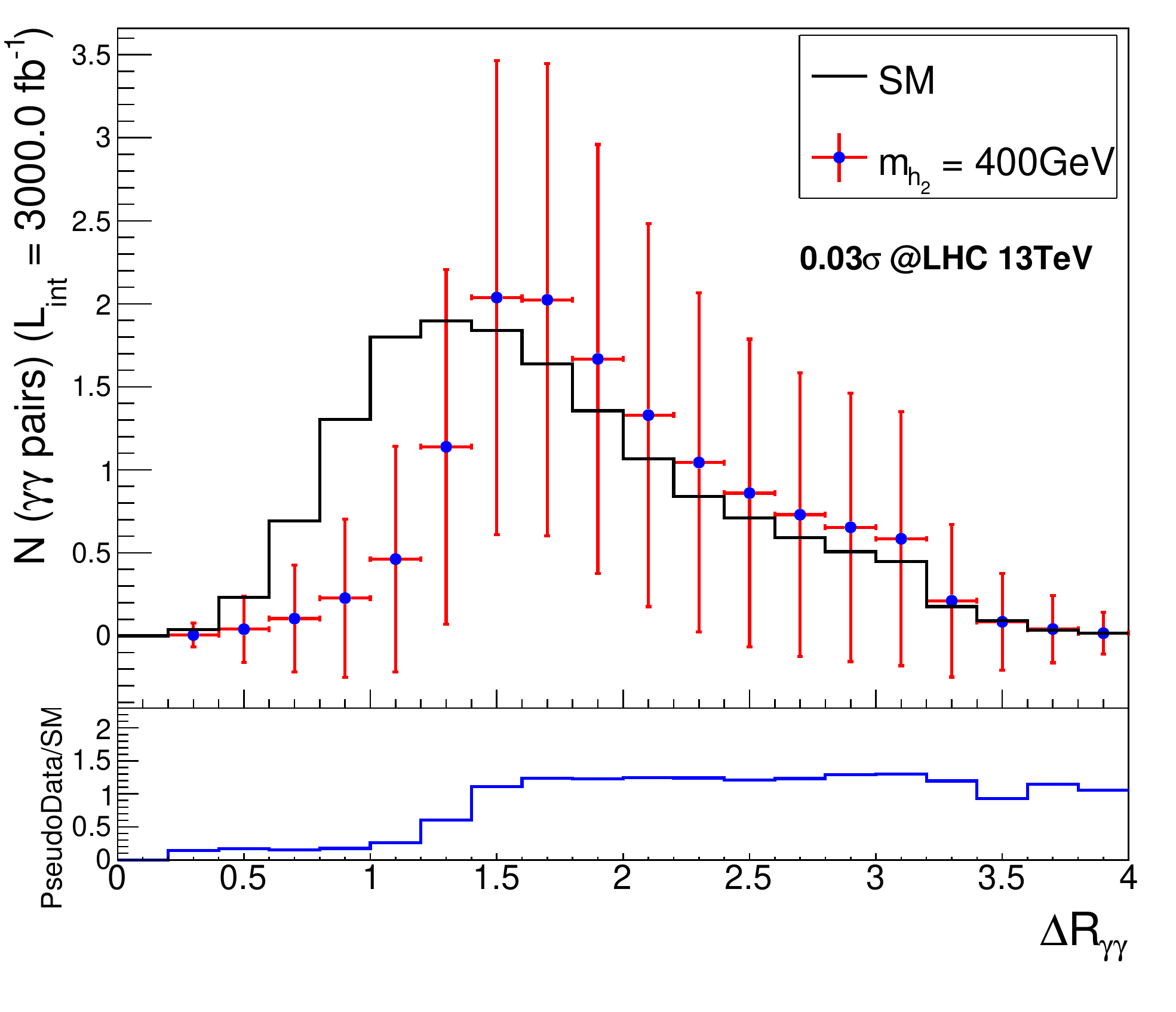}
        \caption{}
        \label{dRaa_L3k}
    \end{subfigure}
    \caption{\small Standard deviation $\chi^2$ test for the benchmark point B$'$ (pseudo data) and SM 
    at 13 TeV LHC with an integrated luminosity of ${\cal L}_{\rm int} =3000\,\,{\rm fb}^{-1}$.}
    \label{statistic}   
 \end{figure}
 
  \begin{figure}
    \centering   
    \includegraphics[width=0.6\textwidth]{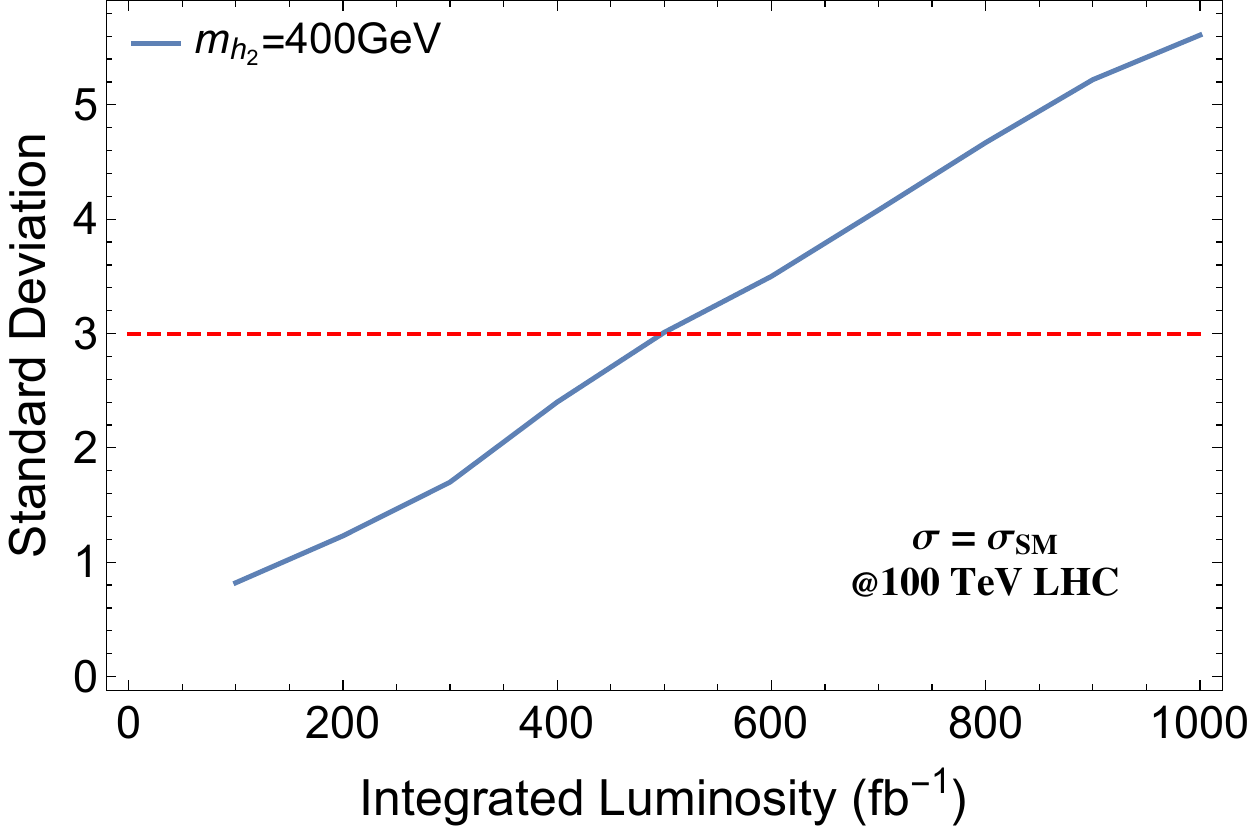}
   \caption{\small The integrated luminosity versus standard deviation $\chi^2$ test for the benchmark point B$'$ and SM at 100 TeV LHC.
   $\sigma = \sigma_{SM}$ means we set the cross section for the process 
   $g g \to h_1 h_1 \to \ga \ga \bar{b} b$ of benchmark point B$'$ to be the same as SM value.}
\label{pairHiggs_stat_Lumi}
 \end{figure}

As mentioned above, the invariant mass distribution of Higgs boson pair peaks at about 400 GeV for the case of the SM, even though there is no resonance. A natural question arises: How well could we tell whether there exists a $h_2$ resonance if the observed total cross section doesn't deviate much from the SM prediction? 
To answer the question, we do shape comparisons.  In Fig.~{\ref{statistic}}a and ~{\ref{statistic}}b,
we perform a standard deviation
$\chi^2$ test for the event distributions in $M_{\gamma\gamma bb}$ and $\Delta R_{\gamma\gamma}$ respectively
at the high luminosity LHC for the SM and a fictitious benchmark point B$'$ (pseudo data)
with $m_{h_2} = 400$ GeV and a total production cross section same as the SM value. 
The error is naively taken as square root of the pseudo data bin content. It turns out that at 13 TeV LHC with ${\cal L}_{\rm int} =3000\,{\rm fb}^{-1}$, we can not tell the difference between new physics signal of the heavy scalar resonance with a mass of 400 GeV and the SM in Higgs boson pair. In particular, it is about 1.31$\sigma$ deviated from the SM in the case of $M_{\ga\ga bb}$ kinematic distribution, and only $0.03\sigma$ in the case of $\Delta R_{\ga \ga}$ distribution. 
However the situation is improved at the future hadron collider.
For illustration, we present the corresponding integrated luminosity versus standard deviation $\chi^2$ test
for the benchmark point B$'$ and SM at 100 TeV LHC. 
The result shown in Fig. {\ref{pairHiggs_stat_Lumi}} indicates that with the integrated luminosity ${\cal L}_{\rm int} \sim 500\,{\rm fb}^{-1}$ at 100 TeV LHC, the signal for a 400 GeV scalar resonance can be distinguished from the SM 
up to $3\sigma$. 
 
\subsection{The  $b\bar{b} b\bar{b}$ Final State Channel}
\label{subsub5.2.3.2}

We select benchmark points E, F, G at which $m_{h_2} = $ 700, 800, 900 GeV respectively 
for studying the kinematic distributions of the $b\bar{b}b\bar{b}$ final state channel. In this case, we follow the event selections used in the ATLAS resolved analysis for $b\bar{b}b\bar{b}$ final state channel \cite{ATLAS:2016ixk}, summarized below:

\begin{itemize}
\item
To be more specific, the events are required to contain at least four $b$-jets with $p_T>30 \, \rm{GeV}$ and $|\eta| < 2.5$. 

\item
Furthermore, we pair up these four $b$-jets to reconstruct two 125 GeV Higgs boson candidates and then impose additional mass-dependent cuts for these two Higgs boson candidates as follows:

\[
\begin{array}{c}
  \frac{360}{m_{4j}/{\rm GeV}} - 0.5 < \Delta R_{jj}^{\rm{lead}} < \frac{653}{m_{4j}/{\rm GeV}} + 0.475 \\
  \frac{235}{m_{4j}/{\rm GeV}} < \Delta R_{jj}^{\rm{subl}} < \frac{875}{m_{4j}/{\rm GeV}} + 0.35   
\end{array}
\Bigg\} \; \;  {\rm{if}} \; m_{4j} < 1250\,  {\rm GeV} \; ,
\]
\[
\begin{array}{c}
  0 < \Delta R_{jj}^{\rm{lead}} < 1 \\
  0 < \Delta R_{jj}^{\rm{subl}} < 1   
\end{array}
\Bigg\} \;\; {\rm{if}} \; m_{4j} > 1250\,  {\rm GeV} \; ,
\]
where $\Delta R_{jj}^{\rm{lead}}$ is the opening angle of two jets which the leading Higgs boson candidate decay into and $\Delta R_{jj}^{\rm{subl}}$ for the sub-leading candidate. Here, the leading Higgs boson candidate refers to the reconstructed Higgs boson that has the larger scalar sum of jet $p_T$. 

\item
Then, an algorithm is applied to select the best pairing of $b$-jets into two Higgs boson candidates as follows. 
A distance $D_{h_1,h_1}$ \cite{ATLAS:2016ixk} is defined as
\begin{equation}
D_{h_1,h_1} = \sqrt{\left( m_{2j}^{\rm{lead}} \right)^2 + \left(m_{2j}^{\rm{subl}} \right)^2} \left| \sin \left(\tan^{-1} \left(\frac{m_{2j}^{\rm{subl}}}{m_{2j}^{\rm{lead}}} \right) - \tan^{-1} \left(\frac{115}{120}\right) \right) \right| \; ,
\end{equation}
which means the distance of the reconstructed pairing's ($m_{2j}^{\rm{lead}}, m_{2j}^{\rm{subl}}$) point to the line connecting the two points (0 GeV, 0 GeV) and (120 GeV, 115 GeV) on the  $m_{2j}^{\rm{lead}} - m_{2j}^{\rm{subl}}$ plane.
Here, $m_{2j}^{\rm{lead/subl}}$ is the mass of the leading/sub-leading Higgs boson candidate and the values of 120 GeV and 115 GeV are basically the centre of signal regions in $m_{2j}^{\rm{lead}}$ and $m_{2j}^{\rm{subl}}$ respectively. 
Among the Higgs boson candidates, the pair that have the minimum distance $D_{h_1h_1}$ is defined to be the two Higgs bosons. 

\item
In additional, the two Higgs boson candidates are required to have the transverse momenta $p^{\rm lead}_T$ and $p_T^{\rm subl}$,  opening angle $\Delta R (h_1,h_1)$ and pseudo-rapidity difference $|\Delta \eta_{h_1h_1}|$ satisfying the following mass-dependent cuts: 
\[
\begin{array}{rl}
  p_T^{\rm{lead}} & > 0.5 \, m_{4j} - 90 \, \rm{GeV} \; ,\\
  p_T^{\rm{subl}} & > 0.33 \, m_{4j} - 70 \, \rm{GeV} \; ,\\     
  \Delta R (h_1,h_1) & > 1.5 \; ,
\end{array}
\]
and
\[
 |\Delta \eta_{h_1h_1}| < \Bigg\{ \begin{array}{c}
  1.1 \hspace{4.7cm} {\rm{if}} \, m_{4j} < 850 \, \rm{GeV} \; , \\
   2 \times 10^{-3}\, \left(m_{4j}/{\rm GeV} \right) - 0.6 \hspace{0.5 cm} {\rm{if}} \, m_{4j} > 850 \, \rm{GeV} \; .
\end{array}
\]

\item
Finally, the mass of Higgs boson candidates must lie in the signal region $X_{h_1,h_1}$ defined by \cite{ATLAS:2016ixk}
\begin{equation}
X_{h_1,h_1} = \sqrt{\left(\frac{m_{2j}^{\rm{lead}} - 120\, {\rm{GeV}}}{0.1\, m_{2j}^{\rm{lead}}}\right)^2 + \left(\frac{m_{2j}^{\rm{subl}} - 115\, {\rm{GeV}}}{0.1\, m_{2j}^{\rm{subl}}}\right)^2} < 1.6 \; .
\end{equation}

\end{itemize}

\begin{figure}[hbtp!]
    \centering   
    \begin{subfigure}{0.49\textwidth}
        \includegraphics[width=\textwidth]{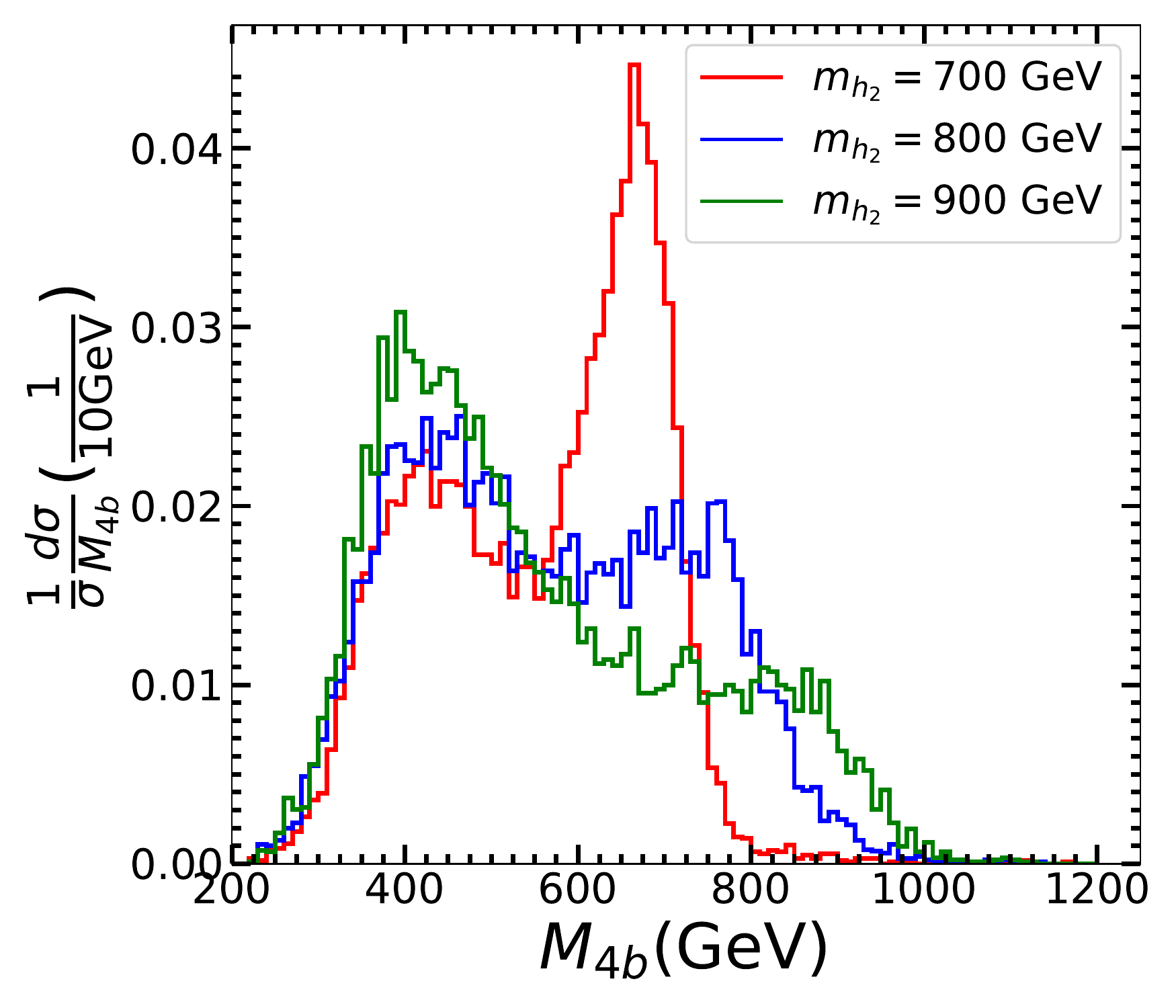}
        \caption{}
        \label{pairHiggs_M4b}
    \end{subfigure}
    \begin{subfigure}{0.49\textwidth}
        \includegraphics[width=\textwidth]{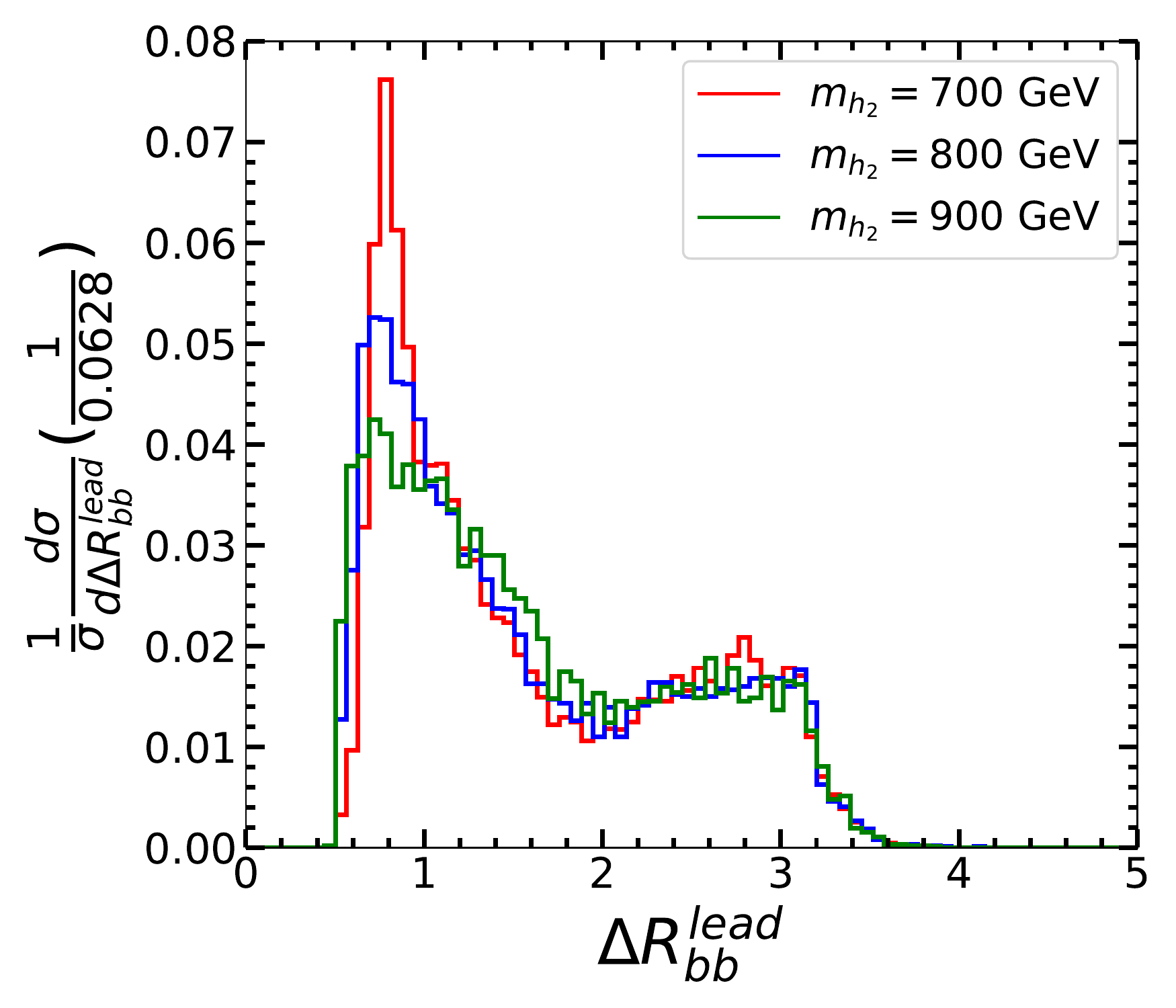}
        \caption{}
        \label{pairHiggs_dRbb1_4b}
    \end{subfigure}
    \begin{subfigure}{0.49\textwidth}
        \includegraphics[width=\textwidth]{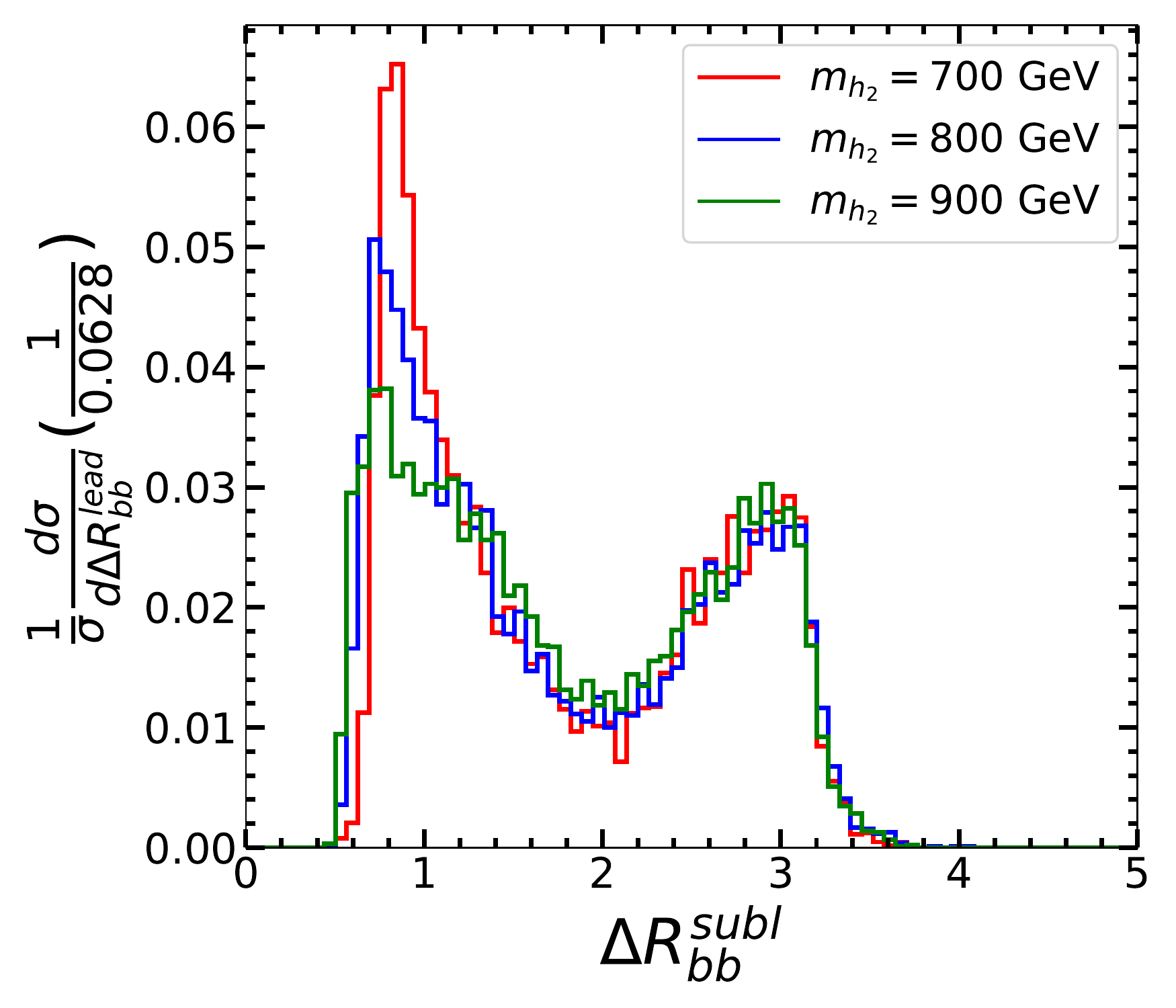}
        \caption{}
        \label{pairHiggs_dRbb2_4b}
     \end{subfigure}
     \caption{\small The kinematic distributions of the $b\bar{b}b\bar{b}$ channel for a) invariant mass of $b\bar{b}b\bar{b}$ $M_{4b}$, 
     b) opening angle $\Delta R_{bb}^{\rm lead}$ of 2 $b$-jets associated with leading Higgs boson candidate and 
     c) opening angle $\Delta R_{bb}^{\rm subl}$ of 2 $b$-jets associated with sub-leading Higgs boson candidate 
     with $m_{h_2} = $ 700, 800, 900 GeV  
     at $\sqrt{s} = 13 \,\rm{TeV}$ for ATLAS detectors.}
     \label{pairHiggs_4b}
\end{figure}

In Fig.~\ref{pairHiggs_4b}, we present the kinematic distributions of the $b\bar{b}b\bar{b}$ channel for a) invariant mass of $b\bar{b}b\bar{b}$, b) opening angle of 2 $b$-jets associated with leading Higgs boson candidate $\Delta R_{bb}^{\rm lead}$ and c) 2 $b$-jets associated with sub-leading Higgs boson candidate $\Delta R_{bb}^{\rm subl}$, with $m_{h_2} = $ 700, 800, 900 GeV at $\sqrt{s} = 13 \,\rm{TeV}$ for ATLAS detectors. We note that these 3 benchmark points E, F, G selected to study this $b\bar{b}b\bar{b}$ final state channel have the same production cross sections and are about twice the SM value. 
In Fig. {\ref{pairHiggs_M4b}}, one can observe that the non-resonant contributions, peaked around 400 GeV, become dominant 
in the benchmark points F and G with $m_{h_2} = $ 800 and 900 GeV respectively, while the benchmark point E with $m_{h_2} = $ 700 GeV 
represents a more dominant contribution from the resonant process. 
Next, in both Figs. {\ref{pairHiggs_dRbb1_4b}} and {\ref{pairHiggs_dRbb2_4b}}, the $\Delta R$ distributions tend to separate into two peaks, 
one locates at $\Delta R \approx 1$, while another at $\Delta R \approx 3$. This behaviour is expected because of the different contributions from non-resonant and resonant processes to the total production cross section. In particular the peak at about $1$ represents the resonant contribution, while the peak at about 3 represents non-resonant contribution. Therefore, in this case, one may be able to use opening angle $\Delta R$ to separate the non-resonant and resonant contributions.
We note that the $\Delta R^{\rm lead}_{bb}$ distribution is more preferable to have a peak located at $\sim 1$, 
due to its more energetic parent Higgs boson.

\section{Conclusion \label{concl}}

Studying Higgs boson pair production is an important way to probe for the Higgs boson self-coupling, one of the important properties of the Higgs boson. 
It is also an interesting channel to search for new physics beyond the SM, especially for models that involve a new resonance decaying into a Higgs boson pair.  
We have studied this process in the G2HDM, a model that promotes the discrete $Z_2$ symmetry ensuring the stability of DM in IHDM to a local gauge symmetry by embedding the two Higgs doublets into the fundamental representation of the new $SU(2)_H$ gauge group. 
One interesting feature in G2HDM is that the SM-like Higgs boson  is a linear combination of the $SU(2)_L$ doublet and the $SU(2)_H$ doublet and triplet scalars, which leads to the deviations in the interactions of Higgs boson to not only other particles but also itself.
In particular, the Higgs boson self-coupling could become negative, which will result in a constructive interference between the box and the triangle Feynman diagrams. Moreover, the existence of the heavy scalar bosons in G2HDM, which will decay into two 125 GeV Higgs bosons, can significantly enhance the production cross section of Higgs boson pair at the LHC. 

We have taken into account theoretical and experimental constraints on the parameter space of the model. The DM constraints,
including DM relic density and direct detection searches, have also been imposed. 
We find out that the Higgs boson trilinear coupling is stringently constrained by the DM relic density and direct searches. 
In particular, the $\lambda_{h_1 h_1 h_1}$, which could vary in the range of $[-29,32]$ before the DM constraints applied, shrinks to be in the range of $[-1, 1.3]$. This results in reducing the double Higgs boson production cross section in G2HDM  
from about two orders of magnitude enhancement before applying DM constraints 
to just about one order of magnitude above the SM value.

We have performed a detailed simulation for the two final states of $b \bar b \gamma \gamma$ and 
$b \bar b b \bar b$ by focusing on a few benchmark points of the parameter space in G2HDM.
The G2HDM demonstrates a representative case in which the kinematic distributions of the $b \bar b \gamma \gamma$ 
and $b \bar b b \bar b$ final state channels are significantly altered from the SM at $\sqrt{s}=13\,\rm{TeV}$ 
LHC.
We also present a standard deviation $\chi^2$ test for a benchmark point 
with $m_{h_2} = 400 \, \rm{GeV}$ and the SM. 
It turns out that it is impossible to distinguish a $400$ GeV resonant signal from the SM 
at the high luminosity LHC running at 13 TeV, if the total cross section of such a resonance is closed to the one in the SM.
However at future machines with higher center-of-mass energies and luminosities 
one expects this signal from new scalar resonance
is discernible.

\section*{Acknowledgments}
We would like to acknowledge the support by the National Center for Theoretical Sciences (NCTS).
This work was supported in part by the Ministry of Science and Technology (MoST) of Taiwan under
Grant Numbers 107-2119-M-001-033-, 105-2112-M-003-010-MY3 and 104-2112-M-001-001-MY3.

\end{document}